\documentclass[12pt]{article}
\usepackage{geometry}
\usepackage{graphicx} 
\usepackage{bm}%
\usepackage{subfigure} 
\usepackage{mathrsfs}
\usepackage{amsmath}
\usepackage{amssymb}
\usepackage{extarrows}
\usepackage{hyperref}
\usepackage[natbibapa]{apacite}
\usepackage{natbib}
\usepackage{threeparttable}
\usepackage{floatrow}
\usepackage{multirow}
\usepackage{color}
\usepackage{diagbox}
\usepackage{enumitem}
\usepackage{adjustbox}%
\usepackage{booktabs}
\usepackage[boxed,ruled,commentsnumbered]{algorithm2e}
\usepackage{tikz}
\usepackage{setspace}
\usetikzlibrary{shapes,arrows}
\usepackage{authblk}%
\usepackage{mystyle}
\usepackage[graphicx]{realboxes}
\usepackage{lineno}

\usepackage{rotating,tabularx}

\title{Sequential Gibbs Sampling Algorithm for Cognitive Diagnosis Models with Many Attributes}
 \author[$\dagger$]{Juntao Wang}
 \author[$\dagger$]{Ningzhong Shi}
 \author[*,$\dagger$]{Xue Zhang}
 \author[,$\ddagger$]{Gongjun Xu \footnote{Correspondence should be sent to Xue Zhang E-Mail: \href{mailto:zhangx815@nenu.edu.cn}{zhangx815@nenu.edu.cn}, or Gongjun Xu E-mail: \href{mailto:gongjun@umich.edu}{gongjun@umich.edu}.}}
 \affil[$\dagger$]{Northeast Normal University}
 \affil[$\ddagger$]{University of Michigan, Ann Arbor}
 \date{}
\begin{document}
\maketitle
\begin{abstract}
Cognitive diagnosis models (CDMs) are useful statistical tools to provide rich information relevant for intervention and learning. 
As a popular approach to estimate and make inference of CDMs, the Markov chain Monte Carlo (MCMC) algorithm is widely used in practice. However, when the number of attributes, $K$, is large, the existing MCMC algorithm may become time-consuming, due to the fact that $O(2^K)$ calculations are usually needed in the process of MCMC sampling to get the conditional distribution for each attribute profile. 
To overcome this computational issue, motivated by \cite{culpepper2018improved},
we propose a computationally efficient sequential Gibbs sampling method, which needs $O(K)$ calculations to sample each attribute profile.
We use simulation and real data examples to show the good finite-sample performance of the proposed sequential Gibbs sampling, and its advantage over existing methods.

\bigskip
\noindent Key words: cognitive diagnosis model, Markov chain Monte Carlo, sequential Gibbs sampling.
\end{abstract} 

\newpage 
\section{Introduction}

In recent years, cognitive diagnosis models (CDMs) have gained great achievements in educational and psychological assessments, where latent binary random vectors are often assumed to represent the presence or absence of multiple fine-grained skills or attributes. 
The CDMs can be viewed as a family of restricted latent class models, with the goal of achieving personalized diagnostic classification.
Compared with the Item Response Theory (IRT) models, the CDMs can provide more informative feedbacks on attribute profiles and allow for the design of more effective intervention strategies \citep{templin2010diagnostic}. 

Many CDMs have been proposed in the literature. An incomplete list contains the Deterministic Input, Noisy “And” gate and Noisy Inputs, Deterministic “And” gate models \citep[DINA and NIDA;][]{Haertel1989RestrictedLCM,junker2001cognitive}, the reduced version of the Reparameterized Unified Model \citep[rRUM;][]{hartz2002bayesian, templin2010diagnostic}, the Deterministic Input, Noisy “Or” gate and Noisy Inputs, Deterministic “Or” gate models \citep[DINO and NIDO;][]{Templin_2006},
 the general diagnostic model \cite[GDM;][]{von2005general},  the log-linear cognitive diagnosis model \citep[LCDM;][]{henson2009defining}, and the generalized DINA model \citep[GDINA;][]{de2011generalized}.

To estimate the CDM parameters and perform classification of examinees, 
the Bayesian MCMC method is one popular approach, as it will not only provide the point estimation but also the whole posterior distributional information for statistical inferences. 
In the Bayesian framework, the MCMC algorithm is used to generate the unique stationary distribution that weakly converges to the true target distribution of parameters of interest.
The MCMC algorithm provides a useful tool to solve many complicated problems in statistics and psychometrics.
In the CDM literature, the Bayesian MCMC estimation of CDMs has also been studied. 
For instance, 
Under the confirmatory setting with the $Q$-matrix prespecified,
\cite{culpepper2015bayesian} proposed an efficient Gibbs sampling for the DINA model, in which all parameters were sampled from their full conditional distributions.
\cite{Chung2014Estimating,CHUNG2019102275} estimated the DINA and rRUM models in the Bayesian framework using a Gibbs sampling algorithm. 
   
\cite{culpepper2018improved}  further proposed a Bayesian sequential Gibbs sampler of the rRUM which samples each latent attribute sequentially from the corresponding conditional Bernoulli distribution. 
In addition, the software ``JAGS" has also been used to fit many common CDMs \citep[e.g.][]{zhan2019using}.
In the exploratory CDM setting, Bayesian method has also been used to estimate the model parameters and the $Q$-matrix jointly under identifiability conditions. For instance,  \cite{Chen2017Bayesian} proposed an easily implemented  MCMC algorithm  through a data augmentation strategy and  item parameter reparameterizations. Following \cite{culpepper2018improved},  
\cite{Culpepperandchen2019Development} proposed a similar sequential sampler   to estimate the $Q$ matirx under the exploratory rRUM. 

In modern psychological, educational and medical applications of CDMs, large-scale data, with large numbers of manifest attributes of interest (denoted by $K$), are often collected. In many applications, the number of the corresponding latent classes $2^K$ could become comparable or even larger than the number of examinees $N$. Examples with large number of latent classes can be found in educational assessment \citep{lee2011cognitive} and the medical diagnosis \citep{wu2016nested}.
The increasing dimension of attributes and items often causes high computational cost and therefore introduces new challenges for the estimation and inference of the CDMs.
 
In this paper, we focus on improving the MCMC with the Gibbs sampling in the setting of many latent attributes  under the cofirmatory CDMs with the $Q$-matrix prespecified. 
Existing MCMC algorithms often directly sample from the posterior distribution of each latent attribute profile \citep[see][]{zhan2019using}, with the whole attribute profile treated as one random sample from a categorical distribution with $2^K$ different categories. Therefore, in order to sample one attribute profile, it is needed to evaluate $2^K$ posterior probabilities of each possible profile candidate. 
The corresponding computational overhead for sampling each individual's attribute profile is of the order $O(2^K)$. 
For a large $K$, this would lead to a significant computational burden and also affect the convergence of the MCMC algorithm. 
 Alternatively,  \cite{ZHAN2015Multidimensional} proposed to model the attributes as independent variables by introducing an independent Bernoulli prior for each attribute, the corresponding sampling method is named as the independent Gibbs sampling in this article. Without modeling the dependence among the attributes, the computation cost of sampling each attribute profile in the independent Gibbs Sampling is then reduced to $O(K)$, however, the independent assumption of attributes is often too strong to satisfy in practice.


 
 Since the computational difficulty for large $K$ mainly arises from the sampling of the attribute profiles, we follow 
  the novel idea of  sequential Gibbs sampler proposed in \cite{culpepper2018improved} to  develop an efficient sequential Gibbs sampling method.
   This work extends \cite{culpepper2018improved}, which focuses on the rRUM model, to  the more general GDINA model under the high-dimensional setting with many attributes; such a high-dimensional setting arises in many applications but the related estimation challenge has not been addressed.
 Following  \cite{culpepper2018improved}, the sequential sampler  samples each attribute separately instead of sampling the attribute profile as a whole, and consequently, the computational overheads of sampling attribute profiles is greatly reduced from $O(2^K)$ to $O(K)$. For a large $K$, the improvement is especially significant as shown in the simulation studies.

The rest of the paper is organized as follows. In Section \ref{2}, we give an overview on the CDMs and a Bayesian formulation for the estimation. Section \ref{3} introduces the proposed sequential Gibbs sampling, with a focus on the estimation of the GDINA model as a general version of CDMs. 
The simulations and real data analyses are shown in Section \ref{5} and Section \ref{6}, respectively. A discussion is given in Section \ref{7}. 
The supplementary materials include more details for the proposed algorithm. 
Source code of the proposed method will be made publicly available upon the acceptance of this work. 

\section{Bayesian GDINA Model} \label{2}

This section focus on the GDINA model as the general framework for CDMs, which include many CDMs as special cases such as DINA, DINO, and Reduced RUM \cite[]{junker2001cognitive,hartz2002bayesian,templin2010diagnostic}.
We present the formulation for the Bayesian GDINA model, which contains the model setup, item parameter priors and population parameter prior. 

\subsection{The DINA and GDINA Models}
In CDMs, the examinee's responses depend on his/her latent attribute profile which is denoted by a $K$-dimensional vector $\bm \alpha=(\alpha_{1},\alpha_{2},\cdots,\alpha_{K})'$, where the superscript $'$ denotes transpose. Each attribute $\alpha_k$ indicates the mastery of attribute $k = 1, \cdots, K$, and there are in total $C=2^K$ latent classes $\bm \alpha \in \{0,1\}^K$. Let the binary vector $\bm Y=(Y_1,\cdots,Y_J)$ represent an examinee's responses to $J$ items. Both $\bm \alpha$ and $\bm Y$ are examinee-specific; a particular examinee $i$'s attribute profile and responses are denoted by $\bm \alpha_i$ and $\bm Y_i$ for $i = 1,\cdots,N$. The $N$ examinees' attribute profiles are random samples from a population distribution with the probability: $\pi_{\bm \alpha}=P(\bm \alpha_i=\bm \alpha),$ where $\sum\limits_{\bm \alpha} \pi_{\bm\alpha}=1,\ 0\leq \pi_{\bm\alpha} <1$. Thus, the population distribution of attribute profiles is characterized by the vector $\bm\pi=(\pi_{\bm\alpha},\bm \alpha \in \{0,1\}^K)'$. 
For notational convenience, for $\bm \alpha=(\alpha_{1},\alpha_{2},\cdots,\alpha_{K})'$, we will also write $\pi_{\bm\alpha}$ as $\pi_c$ with $c=1+\sum_{k=1}^K \alpha_{k} 2^{k-1}$. Note that the two representations $\bm\pi=(\pi_{\bm\alpha},\bm \alpha \in \{0,1\}^K)'$ and $\bm\pi=(\pi_{c},c= 1,\cdots ,C)'$ are equivalent. 


The binary $Q$-matrix \citep{tatsuoka1983rule} is a key component for CDMs. For each pair of $j$ and $k$, $q_{jk}=1$ indicates attribute $k$ is required by item $j$, otherwise $q_{jk}=0$, for $j = 1, \cdots, J$. 
Particularly, the $j$th row vector $\bm q_j$ of the $Q$-matrix corresponds to the attributes required by item $j$.

The DINA model \citep{Haertel1989RestrictedLCM,junker2001cognitive} is one of, if not the simplest, consequently most restrictive, interpretable CDMs available for dichotomously scored tests. For a specific examinee with an attribute profile $\bm \alpha$, we can define the ideal response $\eta(\bm\alpha, \bm q_j)$ to item $j$ relying on $\bm\alpha$ and $\bm q_j$ as 
\begin{equation} \label{}
\eta(\bm\alpha, \bm q_j) = \prod_{k=1}^K \alpha_{k}^{q_{jk}}. \label{www}
\end{equation}
For brevity, given examinee $i$'s attribute profile $\bm\alpha_i$, the ideal response $\eta(\bm\alpha_i, \bm q_j)$ can also be written as $\eta_{ij}$ if the context permits. The $ \eta_{ij}$ is an indicator of whether examinee $i$ masters all the required attributes for item $j$, which indicates that each item partitions all examinees into two latent groups. Let $g_j = P(Y_{ij}=1|\eta_{ij}=0)$ and $s_j = P(Y_{ij}=0|\eta_{ij}=1)$ be the guessing and slipping parameters, respectively. For examinee $i$ and item $j$, the positive response probability, denoted by $\theta_{j,\bm \alpha_i}= P(Y_{ij}=1| \bm \alpha_i)$, takes the form
\begin{equation} \label{}
\theta_{j,\bm \alpha_i} =g_j^{1-\eta_{ij}}(1-s_j)^{\eta_{ij}}.
\end{equation}

\cite{de2011generalized} proposed a general framework  for CDMs based on the DINA model, called the GDINA model, which characterized more complex relationships between attribute profiles and response data. In the GDINA model, the positive response probability can be decomposed into the sum of the effects due the presence of required attributes and their interactions. We let ${K_j^*}=\sum_{k=1}^Kq_{jk}$ be the number of required attributes by item $j$, which is determined by the $j$th row vector $\bm q_j$ in the $Q$-matrix. For a specific examinee with $\bm \alpha$ and item $j$, we rearrange the structure of attribute profiles, so that the first $K_j^*$ attributes are the attributes required by item $j$. The reduced attribute profile for item $j$ consists of the first $K_j^*$ required attributes denoted by $\bm \alpha_{j}^*=(\alpha^*_{j1},\cdots,\alpha^*_{jK_j^*})'$, $j= 1,\cdots,J$. Similarly to $\bm \alpha$ and $\bm Y$, there also exists the examinee-specific reduced attribute profile, denoted by $\bm \alpha_{ij}^*=(\alpha^*_{ij1},\cdots,\alpha^*_{ijK_j^*})'$. 
Given $\bm \alpha^*_j$, the item response probability of item $j$ is modeled as 
\begin{equation}\label{gdina-eq}
h(\theta_{j,\bm \alpha_j^*})=\lambda_{j0}+\sum_{k=1}^{K^*_j} \lambda_{jk}\alpha^*_{jk}+\sum_{k=1}^{K^*_j-1}\sum_{k'=k+1}^{K^*_j} \lambda_{jkk'}\alpha^*_{jk}\alpha^*_{jk'} +\cdots+\lambda_{j12\cdots{K_j^*}}\prod_{k=1}^{K_j^*}\alpha^*_{jk},
\end{equation}
where
$\theta_{j,\bm \alpha_j^*} = P(Y_{ij}=1 | \bm \alpha^*_j)$ represents the positive response probability of the examinees with the reduced $\bm \alpha^*_j$ to item $j$, $h(\cdot)$ is the link function where usually probit, identity, log and logit links can be employed, $\lambda_{j0}$ is the intercept, $\lambda_{jk}$ is the main effect corresponding to $\alpha^*_{jk}$, $\lambda_{jkk'}$ is the two-way interaction corresponding to $\alpha^*_{jk}$ and $\alpha^*_{jk'}$, $\dots$, $\lambda_{j12 \cdots K_j^*}$ is the $K_j^*$-way interaction corresponding to all required attributes. 
We let $\bm\lambda_j = (\lambda_{j0},\lambda_{j1},\cdots,\lambda_{jK_j^*},\lambda_{j12},\lambda_{j13},\cdots,\lambda_{j12\cdots{K_j^*}})'$ represent the item parameters for item $j$ and $\bm\lambda = (\bm\lambda_1, \cdots, \bm\lambda_J)$ represent the item parameters for all items. The number of item parameters is determined by the structure of $Q$-matrix. Specifically, for item $j$, the number of item parameters is $2^{K_j^*}$.
In this paper, we shall focus on the probit link function, whereas the proposed method can be applied to other link functions as well.

Under the GDINA model, each item $j$ can divide examinees into $2^{K_j^*}$ latent groups. Because $\bm\alpha_{j}^*$ is a sub-vector of $\bm \alpha$, we should notice that $\theta_{j,\bm \alpha}=\theta_{j,\bm \alpha_j^*}$. For equation \eqref{gdina-eq}, we can use the vector-notation to rewrite the positive response probability as follows
 $$h(\theta_{j,\bm \alpha}) = h(\theta_{j,\bm \alpha_j^*})=\bm X_{\bm \alpha_{j}^*}' \bm \lambda_j,$$ 
 where $\bm X_{\bm \alpha_{j}^*} =(1,\alpha^*_{j1},\cdots, \alpha^*_{jK_j^*}, \alpha^*_{j1}\alpha^*_{j2},\alpha^*_{j1}\alpha^*_{j3},\cdots,\prod_{k=1}^{K_j^*} \alpha^*_{jk})'$ denotes a $2^{K_j^*}$-dimensional vector relying on $\bm \alpha_{j}^*$. For a particular examinee $i=1,\cdots,N$, the examinee-specific $\bm X_{\bm \alpha_{j}^*}$ is denoted by $\bm X_{\bm \alpha_{ij}^*}=(1, \alpha^*_{ij1}, \cdots,\alpha^*_{ijK_j^*}, \alpha^*_{ij1}\alpha^*_{ij2},\alpha^*_{ij1}\alpha^*_{ij3},\cdots,\prod_{k=1}^{K_j^*} \alpha^*_{ijk})'$. The GDINA model degenerates to the DINA model by setting all item parameters, except $\lambda_{j0}$ and $\lambda_{j12\cdots{K_j^*}}$, to zero. Then, we can obtain $g_j=h^{-1}(\lambda_{j0})$ and $1-s_j=h^{-1}(\lambda_{j0}+\lambda_{j12\cdots{K_j^*}})$, where $h^{-1}$ is the inverse function of $h$. 

The collection of positive response probabilities is denoted by a ${J\times C}$ matrix $\Theta=(\theta_{j,\bm\alpha})$, which may depend on different forms of item parameters in different CDMs.  Given the response data and all attribute profiles, the conditional likelihood function takes the following form:
\begin{equation} \label{}
p(\bm Y|\bm \alpha,\Theta) = \prod_{i=1} ^N \prod_{j=1} ^J\theta_{j,\bm \alpha_i}^{Y_{ij}}(1-\theta_{j,\bm \alpha_i}) ^{1-Y_{ij} }.
\end{equation}
After integrating attribute profiles, the marginal likelihood function takes the following form:
\begin{equation} \label{}
p(\bm Y|\bm \pi,\Theta) = \prod_{i=1}^N \sum_{c=1}^C\pi_c\prod_{j=1} ^J\theta_{j,\bm \alpha_c} ^{Y_{ij}}(1-\theta_{j,\bm \alpha_c}) ^{1-Y_{ij} }.
\end{equation}

\subsection{Priors of Measurement Models' Parameters}

The population proportion parameter $\bm\pi$ includes the saturated information about the attribute profile distribution in  CDMs.
The Dirichlet distribution is commonly used as a conjugated prior for $\bm\pi$, such as \cite{culpepper2015bayesian},  \cite{culpepper2018improved} and \cite{zhan2019using}. 
The specific form of the Dirichlet prior for $\bm\pi=(\pi_1, \cdots , \pi_C)$ is 
$$ \bm \pi \sim \mathrm{Dirichlet}(\bm\delta).$$
where  $\bm\delta=(\delta,\cdots,\delta_C)'$ represent a $C$-dimensional hyper-parameter vector for $\bm\pi$.

In different CDMs, item parameters will be presented in the different forms, such as $g_j$ and $s_j$  in the DINA model,  and $\bm \lambda_j$  in the GDINA model.
 For the DINA model, independent Beta distributions, $\mathrm{Beta}(a_g,b_g)$ and
$\mathrm{Beta}(a_s,b_s)$, are often used as the priors for guessing and slipping parameters, respectively. We may also constraint $0 \leq g_j<1-s_j\leq1$, to ensure the model identifiability \citep{junker2001cognitive,Chen2017Bayesian,xu2016identifiability,gu2019sufficient}. 
For the GDINA model, 
the normal distributions are often taken as priors for the item parameters $\bm\lambda$ \citep[e.g.,][]{zhan2019using}.  
Specifically,  two types of priors are often chosen:  one is a multivariate normal distribution, $\bm\lambda_j \sim N(\bm \mu_{\lambda_j},\bm \Sigma_{\lambda_j})$, as a general choice; the other is a truncated multivariate normal distribution, $\bm\lambda_j \sim N(\bm \mu_{\lambda_j},\bm \Sigma_{\lambda_j}) \mathcal I_{\{\bm\lambda_j \in \bm T \}}$, which is used to ensure certain monotonicity assumption of the item response function.
Here  $\mathcal I_{\{\cdot \}}$ denotes the   indicator function, 
$\bm T =\{T_1,\cdots,T_m,\cdots, T_{2^{K_j^*}}\}$, and each   $T_m$ represents some pre-specified constraint of the $m$-th element of $\bm \lambda_{j}$. For instance, we may restrict the main effect terms in $\bm \lambda_{j}$ to be positive to ensure the monotonicity assumption.


As discussed in the introduction, 
for a large $K$, 
the existing MCMC algorithms with Dirichlet prior for $\bm\pi$ often suffer from the increasing computational cost of sampling each latent attribute profile $\bm\alpha_i$ from its conditional distribution, which is a categorical distribution with $2^K$ different categories. 
Therefore, it needs to evaluate $2^K$ posterior probabilities of each possible profile candidate to sample each $\bm\alpha_i$, and 
the corresponding computational overhead for sampling  $\bm\alpha_i$ is  $O(2^K)$. 
For a large $K$, this would lead to a significant computational burden and also affect the convergence of the MCMC algorithm. 
	
\section{The Sequential Gibbs Sampling} \label{3}
In this section, we introduce the  sequential Gibbs sampling method, which samples each attribute separately and is computationally efficient for large $K$. 
The sequential Gibbs sampling algorithm will be derived for the GDINA model. 
 It's natural to apply the sequential Gibbs sampling to other CDMs,  and  an example of the DINA model is given in Appendix \ref{SEQDINA}.
\subsection{Motivation}
With the commonly used  Dirichlet prior for $\bm\pi$, many existing Gibbs sampling methods suggest that the full conditional distribution for $\bm \alpha$ takes the following from: 
\begin{equation}
\label{E2}
p(\bm\alpha|*)\propto p(\bm Y|\bm\alpha,\bm\lambda) p(\bm\alpha|\bm\pi),
\end{equation}
where the “$*$” represented all the other parameters and responses. To infer a specific examinee's attribute profile, we need to calculate the posterior probability $p(\bm\alpha_c|*)$ for $c= 1,\cdots,C$  to obtain the posterior distribution. For large $K$, the computation is challenging. Since in this Gibbs sampling method, the whole $\bm\alpha$ should be sampled simultaneously, hereafter this sampling method is referred to as the \textit{simultaneous Gibbs sampling}.

 Following  \cite{culpepper2018improved}, we first describe the sequential sampling method for the attributes.
Let  $\bm\alpha_{\backslash k}$ denote the sub-vector of $\bm\alpha$ excluding the $k$-th attribute. Based on the fact that knowing $\bm\alpha_{\backslash k}$ and $\alpha_{k}$ is equivalent to knowing the attribute profile $\bm\alpha$, it's obvious that $p(\bm\alpha|\bm \pi)=p(\bm\alpha_{\backslash k}, \alpha_{k}|\bm \pi)$. According to Bayes' theorem, given the $\bm\alpha_{\backslash k}$ and $\bm \pi$, the conditional probability of $\alpha_k$ is
\begin{equation}
\label{seq2}
\begin{aligned}
p(\alpha_{k}|\bm\alpha_{\backslash k},\bm \pi)=~&p(\bm\alpha_{\backslash k}, \alpha_{k}|\bm \pi)/p(\bm\alpha_{\backslash k}|\bm \pi)\\
=~&p(\bm\alpha|\bm \pi)/p(\bm\alpha_{\backslash k}|\bm \pi),
\end{aligned}
\end{equation}
Considering Equation (\ref{seq2}), the full conditional distribution for $\alpha_k$ is calculated from 

\begin{equation}\label {E1}
p(\alpha_k|*, \bm\alpha_{\backslash k})\propto p(\bm Y|\bm\alpha,\bm\lambda) p(\alpha_k| \bm\alpha_{\backslash k}, \bm \pi).
\end{equation}
In Equation (\ref{E1}), $p(\bm Y|\bm\alpha,\bm\lambda)$ is the conditional likelihood function.

For the second term on RHS of Equation (\ref{E1}), noticing the binary nature of $\alpha_k\! \in\! \{ {0,1} \}$, we know
that  conditional on $\bm\alpha_{\backslash k}$ and $\bm \pi$,
\begin{equation}
\begin{aligned}
\label{seq2prior}
\alpha_{k}|\bm \alpha_{\backslash k },\bm \pi &\sim \mathrm{Bernoulli}(p_{k\mid \bm \alpha_{\backslash k },\bm \pi})\\
\end{aligned}
\end{equation}
 with 
 $$p_{k\mid \bm \alpha_{\backslash k },\bm \pi}=\frac{\sum_{c=1}^C \pi_c \mathcal I_{\{ \alpha_{ck}=1\cap \bm\alpha_{c\backslash k}=\bm\alpha_{\backslash k}\}}}{ \sum_{c=1}^C \pi_c \mathcal I_{\{ \bm\alpha_{c\backslash k}=\bm\alpha_{\backslash k}\}}},$$
where we use the notation $\bm \alpha_{c \backslash k}$ to represent the $\bm \alpha$ vector corresponding to a general latent class $c$ excluding the $k$-th attribute. 
The above indicator function $\mathcal I_{\{ \alpha_{ck}=1\cap \bm\alpha_{c\backslash k}=\bm\alpha_{\backslash k}\}}$ only selects one component from $\bm \pi$, and the indicator function $I_{\{\bm\alpha_{c\backslash k}=\bm\alpha_{\backslash k}\}}$ can select two components from $\bm \pi$. According to the two indicator functions, we can construct a ratio as $p_{k\mid \bm \alpha_{\backslash k },\bm \pi}$. 
For different CDMs, the Bernoulli distribution form to describe $\alpha_k$ remains unchanged.
Here we can interpret $p_{k\mid \bm \alpha_{\backslash k },\bm \pi}$ as the \textit{``prior" conditional probability} before incorporating any information of the responses. The examinee-specific $p_{k\mid \bm \alpha_{\backslash k },\bm \pi}$ depends on the population parameter $\bm \pi$ and examinee-specific $\bm\alpha_{\backslash k}$; that is, for a specific examinee $i$ with attribute profile $\bm \alpha_{i}$, $p_{k\mid \bm \alpha_{i\backslash k },\bm \pi}$ depends on $\bm \pi$ and $\bm\alpha_{i\backslash k}$, $i=1,\cdots,N$. When there is no ambiguity, we will write $p_{k\mid \bm \alpha_{\backslash k },\bm \pi}$ and $p_{k\mid \bm \alpha_{i\backslash k },\bm \pi}$ as $p_k$ and $p_{ik}$ in the following. 

The Equations (\ref{E1}) and (\ref{seq2prior}) imply a sampling method that can sample the latent attributes sequentially one by one. 
Without loss of generality, the attributes are sampled in an increasing order (i.e., $\alpha_1,\cdots,\alpha_K$ are sampled in turns). 
In Table \ref{tseq2proir}, an example with three attributes is presented to show how Equation (\ref{seq2prior}) works. An 8-dimensional vector $\bm\pi$ (i.e., $K=3$) is used to represent the saturated population information and $\alpha_1$, $\alpha_2$ and $\alpha_3$ are generated in turns. The first two rows show a one-to-one mapping between $\bm\alpha$ and $\bm\pi$. Similar to Gibbs sampling, an initial value of the attribute profile is needed as the starting point. Without loss of generality, let the initial value of $\bm \alpha$ equal to $(000)$. When to sample the first attribute $\alpha_1$, $\alpha_2=\alpha_3=0$ is used in Equation (\ref{seq2prior}), then the first attribute $\alpha_1$ can be drawn from a Bernoulli distribution with $p_1=\frac{\pi_2}{\pi_1+\pi_2}$, which is the prior conditional probability of $\alpha_1=1$ given $\alpha_2=\alpha_3=0$. Assuming the realization of the first attribute $\alpha_1$ is $1$, then we can sample the second attribute $\alpha_2$, conditional on $\alpha_1=1$ and $\alpha_3=0$, from a Bernoulli distribution with $p_2=\frac{\pi_4}{\pi_2+\pi_4}$ in Table 1. Assuming the realization of   $\alpha_2$ is $0$, then we move on to sample $\alpha_3$, conditional on $\alpha_1=1$ and $\alpha_2=0$, from a Bernoulli distribution with $p_3=\frac{\pi_6}{\pi_2+\pi_6}$ in Table 1.
\begin{small}
\begin{table}[htp]
\caption{A sample with three attributes for the conditional Bernoulli distribution}
\label{tseq2proir}
\begin{threeparttable}
\begin{tabular}{ccccccccccc}
\hline
\textbf{$\bm \alpha$}&000&100&010&110&001&101&011&111&$p_k$&Prob\\
\hline
$\bm\pi$&$\pi_1$&$\pi_2$& $ \pi_3$ & $ \pi_4$ & $ \pi_5$ & $ \pi_6$ & $ \pi_7$ &$ \pi_8$ & $-$& $-$\\
$\alpha_{1}=1|$ $\alpha_{2}$=0,$\alpha_{3}$=0,$\bm\pi$ &$\pi_1$ & $\pi_2$ & $ -$ & $-$ & $ -$ & $-$ & $ - $&$ -$&\tiny{$\frac{\pi_2}{\pi_1+\pi_2}$}& $p_1$\\
$\alpha_{2}=0|$ $\alpha_{1}$=1,$\alpha_{3}$=0,$\bm\pi$&$ -$ & $ \pi_2$ & $ -$ & $\pi_4$ & $ -$ & $ -$ & $ - $ &$ - $& \tiny{$\frac{\pi_4}{\pi_2+\pi_4}$}&$1-p_2$\\
$\alpha_{3}=1|$ $\alpha_{1}$=1,$\alpha_{2}$=0,$\bm\pi$&$ -$ & $\pi_2$ & $ -$ & $ -$ & $ - $ & $ \pi_6$ & $ -$ &$ - $& \tiny{$\frac{\pi_6}{\pi_2+\pi_6}$}&$p_3$\\
\hline
\end{tabular}
\begin{tablenotes}
        \footnotesize
        \item Note. The column “$p_k$” represents the conditional probability of $\alpha_k=1$. The column “Prob" is the probability of realization $\alpha_k$ shown in the first column (in this table, the realizations are $\alpha_{1}=1$, $\alpha_{2}=0$ and $\alpha_{3}=1$). 
      \end{tablenotes}
\end{threeparttable}
\end{table}
\end{small}

In both of the sequential and simultaneous sampling methods,
the sampling of attribute profiles  depends on $\bm \pi$ and $\bm \delta$. However,
in the simultaneous Gibbs sampling, each attribute profile is treated as a basic unit, and the joint information $p(\bm \alpha|\bm \pi)$ is used to sample $\bm\alpha$. This method is very slow when $K$ is large. In the sequential Gibbs sampling, each element of the attribute profile is sampled seperately from the conditional Bernoulli distribution of   $\alpha_{k}|\bm\alpha_{\backslash k},\bm \pi$, which would reduce the compuational cost significantly. 

\subsection{Sequential Gibbs Sampling Schedules}

With the above introduced sequential sampling method for attributes, in this section we derive the Gibbs sampling updates for other model parameters. 
To illustrate our method, we shall focus on the GDINA model with a probit link function and use the prior settings introduced in Section 2.2. 


  We will use a data augmentation strategy to derive a closed-form Gibbs sampling method for the item parameters.  Please note that similar sampling methods have been proposed in the CDM literature \citep{chen2020sparse,culpepper2019exploratory,culpepper2019estimating}.
Specifically, we introduce the data augmentation process for the examinee with $\bm \alpha$ to item $j$ as follows $$Z_{j}=\bm X_{\bm \alpha_{j}^*}' \bm \lambda_j+\varepsilon_{j},$$ 
where $\varepsilon_{j}$ follows a standard normal distribution and   $Z_j$ is a latent auxiliary variable. The $Z_j$ is the examinee-specific, and the augmented data of item $j$ for examinee $i$ is denoted by $Z_{ij}$, for $i=1,\cdots,N$. For examinee $i$, the augmented data $Z_{ij}$ is distributed $N(\bm X_{\bm \alpha_{ij}^*}' \bm \lambda_j,1)$, and the item response $Y_{ij}$ is defined as $Y_{ij}=1$ if $Z_j$ is positive, and $Y_{ij}=0$ otherwise.

 With  the introduced augmented data $\bm Z$, the Gibbs sampling needs to  sample from the four full conditional distributions:
$p(\bm Z | \bm Y, \bm\alpha,\bm\pi,\bm\lambda)$,
$p(\bm\lambda| \bm Y,\bm Z , \bm\alpha,\bm\pi)$,
$p(\alpha_{ik} | \bm Y, \bm Z,\bm \alpha_{i\backslash k},\bm\pi,\bm\lambda)$ and
$p(\bm\pi | \bm Y, \bm Z, \bm\alpha,\bm\lambda)$.

\paragraph{Sample Augmented Data.} For examinee $i$ and item $j$, the augmented data is $Z_{ij}$. Conditional on $\bm\alpha$, the distribution of $Z_{ij}$ is independent of the parameter $\bm\pi$, which means the distributions $p(\bm Z | \bm Y, \bm\alpha,\bm\pi,\bm\lambda )$ and $p( \bm Z | \bm Y, \bm\alpha, \bm\lambda)$ are equivalent.

According to the $j$th row vector $\bm q_j$ in the $Q$-matrix, we can get the reduced vector $\bm \alpha_{ij}^*$. Based on $\bm \alpha_{ij}^*$, $\bm \lambda_j$ and $Y_{ij}$, the augmented data $Z_{ij}$ is normally distributed with the mean $\mu_{ij}=X_{\bm \alpha_{ij}^*}' \bm \lambda_j
$ and the variance one. The range of $Z_{ij}$ is determined by $Y_{ij}$; if $Y_{ij}=1,\ Z_{ij}>0$, otherwise $Z_{ij} \leq 0$. The augmented data is generated by the formula%
\begin{equation}
\label{aug}
Z_{ij}|Y_{ij},\bm \alpha_{ij}^*,\bm \lambda_j \sim
\begin{cases}
N(\mu_{ij},1)\mathcal I_{\{z_{ij}>0\}}& Y_{ij}=1\\
N(\mu_{ij},1)\mathcal I_{\{z_{ij}\leq0\}}& Y_{ij}=0\\
\end{cases}.
\end{equation} 

\paragraph{Sample Item Parameters.} In the GDINA model, two considered types of item parameter priors are the multivariate normal distribution and the truncated multivariate normal distribution, which will induce two sampling methods to sample item parameters. 
The truncated prior is suitable for the case we have known some constrains on item parameters. 
The multivariate normal distribution is suitable for the case we don't have additional information about item parameters. 

For the item parameters, the conditional independence implies $p(\bm\lambda| \bm Y,\bm Z , \bm\alpha,\bm\pi)$ and $p(\bm\lambda| \bm Y,\bm Z , \bm\alpha)$ are equivalent. To sample the item parameters for item $j$, the information of all examinees for this item need to be considered. We arrange all examinees' augmented data about item $j$ in a vector $\bm Z_{j}= (Z_{1 j},Z_{2 j},\cdots,Z_{N j})'$. Furthermore, let $\bm X_j =(\bm X_{\bm \alpha_{1j}^*},\bm X_{\bm \alpha_{2j}^*},\cdots,\bm X_{\bm \alpha_{Nj}^*})'$ denote an $N \times 2^{K_j^*}$matrix relying on $\bm \alpha^*_{ij}$ for $i =1,\cdots,N$. 
Given $\bm Z_j$ and $\bm X_j$, a linear regression model is obtained as follows:
$$\bm Z_{j} = \bm X_j \bm \lambda_j + \bm\varepsilon_j,$$
where $\bm\varepsilon_j=(\varepsilon_{1j},\varepsilon_{2j,}\cdots,\varepsilon_{Nj})'$ is a random vector from a standard normal distribution. If there are no constraints on the item parameter $\bm\lambda_j$, which follows a the prior $N(\bm \mu_{\lambda_j},\bm\Sigma_{\lambda_j})$, then we can obtain the full conditional distribution \citep{minka2000bayesian} whose form is shown as
\begin{equation}
\label{closedform1}
\bm \lambda_j|\bm \alpha,\bm Z_{ j} \sim N(\bm {\hat \mu}_{\lambda_j},\bm {\hat \Sigma}_{\lambda_j}),
\end{equation}
where $\bm{\hat \Sigma}_{\lambda_j}^{-1}=\bm \Sigma_{\lambda_j}^{-1} +\bm X_j\bm X_j'$ and $\bm{\hat \mu}_{\lambda_j}=\bm{\hat \Sigma}_{\lambda_j} (\bm X_j '\bm Z_{ j}+ \bm \Sigma_{\lambda_j}^{-1} \bm \mu_{\lambda_j})$. The sampling method using Equation (\ref{closedform1}) is called the sampling without truncation. The specifics of the derivation can be found in the Appendix \ref{colsed form}. 
If the prior of $\bm\lambda_j$ is the truncated distribution $N(\bm \mu_{\lambda_j},\bm \Sigma_{\lambda_j})\mathcal I_{\{\bm\lambda_j \in \bm T \}}$, we can obtain the closed form for $\bm \lambda_j$'s full conditional distribution:
\begin{equation}
\label{closedform2}
\bm \lambda_j|\bm \alpha,\bm Z_{j} \sim N(\bm {\hat \mu}_{\lambda_j},\bm {\hat \Sigma}_{\lambda_j}) \mathcal I_{\{\bm\lambda_j \in \bm T \}}.
\end{equation}
The sampling method using Equation (\ref{closedform2}) is called the sampling with truncation. 
The details about how to sample from the truncated multivariate normal distribution will be discussed in the Appendix \ref{sampling}.

\paragraph{Sample Attribute Profiles.} In the sequential Gibbs sampling, attributes are sampled one by one, instead of the whole attribute profile. For examinee $i$, if the $k$-th attribute $\alpha_{ik}$ isn't required by an item, the value of $\alpha_{ik}$ won't affect the item's likelihood. So when to sample attribute $\alpha_{ik}$, we only need to pay attention to the items requiring the $k$-th attribute. Hence, we define a set $\hat {\Omega}_{k}=\{j\mid q_{jk}=1,j=1,\cdots,J\}$, which represents the items which require attribute $\alpha_k$, and the complementary set of $\hat {\Omega}_{k}$ is defined as $\hat {\Omega}^c_{k}=\{j\mid q_{jk}=0,j=1,\cdots,J\}$. Only the items from $\hat {\Omega}_{k}$ will affect the inference about $\alpha_{k}$.

Assuming item $j$ belongs to $\hat {\Omega}_{k}$ and giving the reduced attribute profile $\bm \alpha_{ij}^*$, the positive response probability $\theta_{j,\bm\alpha_i} = \Phi(\bm X_{\bm \alpha_{ij}^*}' \bm \lambda_j)$. For the specific examinee $i$, the likelihood function for the $k$-th attribute $\alpha_{ik}$ is 
\begin{equation}
\label{gdinaa}
\begin{aligned}
 p(Y_{ij}|\bm\alpha_{ij}^*,\bm \lambda_j)
=~&\Phi(\bm X_{\bm \alpha_{ij}^*}' \bm\lambda_j)^{Y_{ij}}(1-\Phi(\bm X_{\bm \alpha_{ij}^*}' \bm\lambda_j))^{1-Y_{ij}}\\
=~&\Phi(T_0^{ij}+\alpha_{ik}T_1^{ij})^{Y_{ij}}(1-\Phi(T_0^{ij}+\alpha_{ik}T_1^{ij}))^{1-Y_{ij}}\\
=~&[\Phi(T_0^{ij}+T_1^{ij})^{\alpha_{ik}}\Phi(T_0^{ij})^{1-\alpha_{ik}}]^{Y_{ij}} \\
&[(1-\Phi(T_0^{ij}+T_1^{ij}))^{\alpha_{ik}}(1-\Phi(T_0^{ij}))^{1-\alpha_{ik}}]^{1-Y_{ij}},\\
\end{aligned}
\end{equation}
where $\bm X_{\bm \alpha_{ij}^*}' \bm\lambda_j =T_0^{ij}+\alpha_{ik}T_1^{ij} $ with the two terms $T_0^{ij}$ and $T_1^{ij}$     defined as follows. For $\bm X_{\bm \alpha_{ij}^*}' \bm\lambda_j$, the notation $T_0^{ij}$ is the sum of the terms which don't contain $\alpha_{ik}$ and $T_1^{ij}\alpha_{ik}$ is the  sum of the terms related to $\alpha_{ik}$. If examinee $i$ masters attribute $\alpha_k$, the positive response probability is $\Phi(T_0^{ij}+T_1^{ij})$, otherwise, the positive response probability is $\Phi(T_0^{ij})$. Therefore, the positive response probability is $\Phi(T_0^{ij}+T_1^{ij})^{\alpha_{ik}}\Phi(T_0^{ij})^{1-\alpha_{ik}}$ and a similar expression, $(1-\Phi(T_0^{ij}+T_1^{ij}))^{\alpha_{ik}}(1-\Phi(T_0^{ij}))^{1-\alpha_{ik}}$, can be obtained for the negative response. For example, assume that the vector $\bm q_j =(110)$, the third attribute doesn't affect the positive response probability and the likelihood function. In other words, from the responses on this item we can't get any information about the third attribute. We show how to calculate $T_0^{ij}$ and $T_1^{ij}$. When to investigate the first attribute $\alpha_{1}$, the positive response probability is that 
\begin{equation}
\begin{aligned}
 &\Phi(\lambda_{j0} + \lambda_{j1}\alpha_1+\lambda_{j2}\alpha_2+\lambda_{j12}\alpha_1\alpha_2)\\
 =~&\Phi(\lambda_{j0} +\lambda_{j2}\alpha_2+\alpha_1(\lambda_{j1}+\lambda_{j12}\alpha_2)),
\end{aligned}
\end{equation}
then notations $T_0^{ij}=\lambda_{j0} +\lambda_{j2}\alpha_2$ and $T_1^{ij}=\lambda_{j1}+\lambda_{j12}\alpha_2$.

According to Equation (\ref{gdinaa}), it’s obvious that only the items in $\hat {\Omega}_{ik}$ will affect the full conditional distribution for $\alpha_{ik}$. The parameters $\bm\pi$ and $\bm \alpha_{\backslash k}$ are used to calculate the prior conditional probability for $\alpha_{ik}$, with  $p_{ik}$ calculated as in Section \ref{3}. Then the full conditional distribution for $\alpha_{ik}$ is calculated by
\begin{equation}
\begin{aligned}
&p(\alpha_{ik}|\bm Y_{i },\bm\alpha_{i\backslash k},\bm \lambda,\bm\pi) \\
\propto~&\prod_{j \in \hat {\Omega}_{k}} p(Y_{ij}|\bm \alpha_i,\bm \lambda_i) p_{ik}^{\alpha_{ik}} (1-p_{ik})^{1-\alpha_{ik}}\\
=~&\prod_{j\in \hat {\Omega}_{k}} [\Phi(T_0^{ij}+T_1^{ij})^{\alpha_{ik}}\Phi(T_0^{ij})^{1-\alpha_{ik}}]^{Y_{ij}}\\
 &[(1-\Phi(T_0^{ij}+T_1^{ij}))^{\alpha_{ik}}(1-\Phi(T_0^{ij}))^{1-\alpha_{ik}}]^{1-Y_{ij}} p_{ik}^{\alpha_{ik}} (1-p_{ik})^{1-\alpha_{ik}}\\
 =~&\left[\prod_{j \in \hat {\Omega}_{k}}\Phi(T_0^{ij}+T_1^{ij})^{Y_{ij}} (1-\Phi(T_0^{ij} +T_1^{ij}))^{1-Y_{ij}} p_{ik}\right]^{\alpha_{ik}}\\
&\left[\prod_{j\in \hat {\Omega}_{k}} \Phi(T_0^{ij})^{Y_{ij}} (1-\Phi(T_0^{ij}))^{1-Y_{ij}} (1-p_{ik})\right]^{1-\alpha_{ik}}
\end{aligned}
\end{equation}
Hence, the full conditional distribution for $\alpha_{ik}$ is $\mathrm{Bernoulli}(\hat p_{ik})$, where the value of $\hat p_{ik}$ is given by 
\begin{small}
\begin{equation}\label{sampA}
\frac {\prod_{j\in \hat {\Omega}_{k}} \Phi(T_0^{ij}+T_1^{ij})^{Y_{ij}} (1-\Phi(T_0^{ij} +T_1^{ij}))^{1-Y_{ij}} p_{ik}} {\prod_{j \in \hat {\Omega}_{k}}\Phi(T_0^{ij}+T_1^{ij})^{Y_{ij}} (1-\Phi(T_0^{ij} +T_1^{ij}))^{1-Y_{ij}} p_{ik}+\prod_{j \in \hat {\Omega}_{k}} \Phi(T_0^{ij})^{Y_{ij}} (1-\Phi(T_0^{ij}))^{1-Y_{ij}} (1-p_{ik})}.
\end{equation}
\end{small}

\paragraph{Sample the Population Parameter.} The population parameter $\bm \pi $ is a $C$-dimensional vector, whose prior is $\mathrm{Dirichlet}(\bm\delta)$. Given $\bm \alpha$, we can calculate the number of examinees within the latent class $c$, $N_c=\sum_{i=1}^N I_{\{\bm\alpha_i=\bm\alpha_c\}}$, and the vector $\bm N= (N_1,N_2,\cdots,N_C)'$. From the conditional indepdendence, we know  $p(\bm\pi | \bm Y, \bm Z, \bm\alpha,\bm\lambda ;\bm\delta)$ and $p(\bm\pi | \bm\alpha;\bm\delta)$ are equivalent. And we can write the posterior of $\bm\pi$ as 
\begin{equation}  \label{samppai1}
\bm\pi\mid \bm\alpha;\bm\delta \sim\mathrm{Dirichlet}(\bm\delta+\bm N).
\end{equation} 

We summarize the sequential Gibbs sampling for the GDINA model in Algorithm \ref{A1}. The sequential sampling method can be straightforwardly applied to other CDMs as well, and the DINA example is illustrated in the Appendix. 

 \begin{algorithm}[htp]
 \label{A1}
            \caption{Sequential Gibbs Sampling for GDINA models}
            \KwIn{Initialize $\bm\lambda^{(0)},\bm\alpha^{(0)},\bm\pi^{(0)},\bm Y,m=0, M$ and specify priors.}
            \KwOut{Markov chains of $\bm\lambda^{},\bm\alpha,\bm\pi$. }

            \While{$m < M$}{
            Generate the augmented data from Equation (\ref{aug}).\\
            Sample item parameters from Equation (\ref{closedform1}) or (\ref{closedform2}).  \\
            Sample attribute profiles from Equation (\ref{sampA}). \\
            Sample the population parameter from Equation (\ref{samppai1}).\\
            Set $m= m+1$
            }
           \end{algorithm}

\section{Simulation Studies} \label{5}

In this section, the simultaneous Gibbs sampling, independent Gibbs sampling \citep{ZHAN2015Multidimensional}\footnote{ For the independent Gibbs sampling, we use the independent Bernoulli prior that $\alpha_{nk}\sim \mathrm{Bernoulli}(p_{ik})$ and $p_{ik} \sim \mathrm{Beta}(1,1)$, where $i$ and $k$ indicate examinee and attribute, respectively. }
, and sequential Gibbs sampling are used to estimate parameters in the DINA and GDINA models. The simulation studies intend to implement on different settings of $K$. However, for large $K$, the simultaneous Gibbs sampling methods doesn't work due to the high computational cost, so only the results of independent and sequential Gibbs sampling are shown. To show that the difference among these methods is purely caused by the difference among the sampling techniques rather than the software, we code and compile  all these three methods by ourselves. The computation of the simulation study is implemented by Dell XPS with 3.0 GHz Intel Core i7-9700, 24 GB RAM.

The statistical software JAGS \citep[Just Another Gibbs Sampling;][]{plummer2003jags}, as the off-the-shelf sampling method, is also used to implement  the simultaneous Gibbs sampling method for the DINA and GDINA models. 
The JAGS is similar to WinBUGS \citep{lunn2000winbugs} and OpenBUGS \citep{foulley2010modelling}. \cite{zhan2019using} showed  how to implement the DINA and linear logistics models \citep[LLM, see][]{maris1999estimating} by JAGS. 
We use JAGS to analyze the DINA and GDINA models. When using JAGS, the initial values of all parameters are generated by the default way within JAGS. 
Under the DINA model, another simultaneous sampler, the R package “dina” \citep{culpepper2015bayesian}, is also used to estimate the model. Our simulation results indicate that “dina” is faster than JAGS \footnote{ Particularly, using “dina” to run the chains with 2000 iterations for the DINA settings in our simulation with $\{N = 2000, K = 3\}, \{N = 2000, K = 5\},$ and $\{N = 2000, K = 7\}$ needs about 20, 82, and 344 seconds, which are faster than JAGS but  slower than the proposed method.} 
and the parameter estimates of “dina” are similar to JAGS. 
As the R package ``dina" can’t handle the GDINA model, the detailed results of ``dina" are not shown.


\subsection{Simulation Design}\label {6.1}

\hangafter 1
\hangindent 0em
The attribute profiles are generated from the two following structures.
\paragraph{Uniform Structure.} The uniform structure assumes that all latent classes share the same probability.
\paragraph{Correlated Structure.} \cite{chiu2009cluster} proposed a correlated structure for attribute profiles, which can be viewed as a special case of the higher-order attribute structure. For each examinee, the $K$-dimensional vector $\bm \theta = (\theta_1, \cdots ,\theta_K)$ follows a multivariate normal distribution $N(\bm 0,\bm \Sigma)$, where the covariance matrix $\bm \Sigma$ has a common correlation $\rho$ as follows
$$\begin{pmatrix}
1&& \rho\\
&\ddots&\\
\rho&&1\\
\end{pmatrix},$$
then attributes are determined by 
\begin{equation}
\alpha_k=\left\{
\begin{aligned}{}
1 &\ &\mathrm{if} \ \theta_k > 0,\\
0 &\ &\mathrm{otherwise}.
\end{aligned}
\right.
\end{equation}
\cite{chen2015statistical} also called this situation as “Dependent Attributes”.


For the DINA and GDINA models, the generation methods of item parameters need to be introduced separately. For the DINA model, we set the guessing and slipping parameters to 0.2. For the GDINA model, another equivalent notation is introduced to make the description of item parameters clear. For item $j$, let $\lambda_j^{(0)}$ and $\lambda_j^{(w)}$ denote the intercept parameter and $w$-way interaction parameter, respectively. 
We generate  $\bm\lambda_j$ from a multivariate normal distribution with a diagonal covariance matrix.
In particular, the distribution to generate $\bm\lambda_j$ is specified as:
\begin{equation}\label{GDTP}
\lambda_j^{(w)} \sim
\begin{cases}
N(-1.2,0.4^2), \quad \ \ w=0,\\
N(0.9,0.3^2)/w^2,\quad w = 1,\cdots,K_j^*.
\end{cases}
\end{equation}
This generation method of item parameters indicates the same-way interactions share similar properties (i.e., the same distribution). 


Based on the model identifiability and generic identifiability restrictions \citep{xu2016identifiability,xu2017identifiability,gu2018partial,chen2020sparse}, the $Q$-matrix has this form
$$Q'=\begin{pmatrix}
I_K,I_K,\widetilde Q_2',\widetilde Q_3', \widetilde Q' \\
\end{pmatrix},$$
where each item in $\widetilde Q_2$ and $\widetilde Q_3$ requires two and three attributes, respectively. In addition, we randomly sample non-zero $\bm q$-vectors  which require three or fewer attributes to fill $\widetilde Q$. Given the attribute profiles, the item parameters, the $Q$-matrix, and the response data can be generated. For different $K$, the $Q$-matrices are fixed and shown in Appendix \ref{Qmatrix}.


The sampling methods are compared from three aspects: speed, parameter estimation accuracy and classification accuracy. 
The running times of the sampling methods are used to reflect the speed, and the bias, root mean squared error (RMSE) and mean squared error (MSE) are used to evaluate the accuracy. 
The average bias, RMSE and MSE, denoted by $\overline {\mathrm{Bias}}$, $\overline {\mathrm{RMSE}}$ and $\overline {\mathrm{MSE}}$, are computed for each type parameter, according to
$\mathrm{\overline {Bias}_{\phi}} = \frac{1}{H}\sum_{h=1}^H \frac{1}{R} \sum_{r=1}^R (\hat\phi_h^r-\phi_h)$, $\mathrm{\overline {RMSE}_{\phi}} = \frac{1}{H}\sum_{h=1}^H  \sqrt{\frac{1}{R} \sum_{r=1}^R (\hat\phi_h^r-\phi_h)^2}$ and $\mathrm{\overline {MSE}_{\phi}} = \frac{1}{H}\sum_{h=1}^H  {\frac{1}{R} \sum_{r=1}^R (\hat\phi_h^r-\phi_h)^2}$, where $\hat\phi_h^r$ denotes the estimation from $r$-th replication of a parameter, $\phi_h$ denotes the true value, and $R$ denotes the number of replications (i.e., for guessing and slipping parameters, $H=J$; for the item parameters in the GDINA model, $H$ is the total number of item parameters denoted by ${\#\{\bm\lambda\}}$; for population parameters, $H=C$). The subscript $\phi$ of the indices is used to discriminate the types of parameters. 
On the other hand, two widely used indices, attribute-wise agreement rate (AAR) and pattern-wise agreement rate (PAR), can be used to examine the classification accuracy. The $\mathrm{AAR} =  {\sum_{r=1}^R \sum_{i=1}^N \sum_{k=1}^K \mathcal I \{\hat\alpha_{ik}^r = \alpha_{ik}\}}/{RNK}$ can be used for arbitrary $K$. 
The $\mathrm{PAR} = {\sum_{r=1}^R \sum_{i=1}^N \mathcal I \{\sum_{k=1}^K | \hat\alpha_{ik}^r - \alpha_{ik} | =0 \}}/{RN}$, however, is usually too small to provide valid information about the classification accuracy when $K$ is large. Hence, we define a more practical measure, $\mathrm{PAR}n =  {\sum_{r=1}^R \sum_{i=1}^N \mathcal I \{\sum_{k=1}^K | \hat\alpha_{ik}^r - \alpha_{ik} | \leq n \}}/{RN}$, which denotes that, for one pattern, up to $n$ misestimated attributes can be tolerated. When $n$ is a positive integer, $\mathrm{PAR}n$ is a relaxation of PAR (i.e., $\mathrm{PAR}0$).

For any estimation of the parameter $\bm \pi$, $\mathrm{\overline {Bias}}_{\bm \pi}=0$ always holds. So the maximum norm is used to replace the bias to evaluate the performance of population parameter estimations. When the true value and estimation of population parameter are $\bm\pi$ and $\hat{\bm{\pi}}^{(r)}$, the maximum norm of difference is $\Vert \bm\pi-\hat{\bm{\pi}}^{(r)} \Vert_\infty=\max_c| \pi_c -\hat\pi_c|$. The maximum norm measures the maximum of absolute deviance. If the estimation repeats $R$ times, the average maximum norm ($\overline {\mathrm{MN}}$) is 
 \begin{equation}
 \nonumber
\begin{aligned}{}
\mathrm{\overline {MN}}_{\bm \pi} = \frac{\sum_{r=1}^R \Vert \bm\pi-\hat{\bm{\pi}}^{(r)} \Vert_\infty} {R}. \\
\end{aligned}
\end{equation}


Table \ref{simulation settings} summaries the simulation study basic settings: sample sizes $N=1000$ and $2000$; the number of the items $J=30$; attribute structures uniform, correlated structures with two correlation levels $\rho= 0.3$ and $0.7$. We call the cases $K=3$ and 5 as the low dimension cases, where the simultaneous and sequential Gibbs sampling are conducted. The cases $K=7$ and 15 are named as the high dimension cases, only the sequential Gibbs sampling is performed. For each particular case,   25 independent response datasets are generated. 

\begin{table}[htp]
\caption{The Settings for Simulation Studies}
\begin{center}
\begin{spacing}{1}
\begin{threeparttable}
\begin{tabular}{lp{1cm}<{\centering}p{1cm}<{\centering}p{1cm}<{\centering}p{1cm}<{\centering}p{1cm}<{\centering}p{1cm}<{\centering}p{1cm}<{\centering}}
\hline
\hline
Examinee Sample sizes & \multicolumn{7}{c}{$N$ = 1000 and 2000}\\
Number of Items &  \multicolumn{7}{c}{$J$ = 30}\\
Attribute Structure &\multicolumn{7}{c}{uniform or correlations $\rho=0.3$ and 0.7}\\
Replications & \multicolumn{7}{c}{$R$ = 25}\\
\hline
\hline
&\multicolumn{3}{c}{$K=3,5$}&\multicolumn{4}{c}{$K=7,15$}\\\cmidrule(lr){2-4}\cmidrule(lr){5-8}
Method&Sim&Seq&{Ind}&S&M&L&{Ind}\\
\hline
\hline
&\multicolumn{3}{c}{DINA}&\multicolumn{4}{c}{GDINA}\\\cmidrule(l){2-4}\cmidrule(lr){5-8}
Chain Length & \multicolumn{3}{c}{2000} &\multicolumn{4}{c}{3000}  \\
Burn-in&\multicolumn{3}{c}{1000} &\multicolumn{4}{c}{2000}  \\
\hline
\hline
\end{tabular}
\begin{tablenotes}
        \footnotesize
        \item \textbf{Note}. The column {``Sim", ``Seq" and ``Ind" represent the simultaneous, sequential and independent }Gibbs samplings, respectively.
        \end{tablenotes}
      \end{threeparttable}
\end{spacing}
\end{center}
\label{simulation settings}
\end{table}%


\begin{figure}
    \centering
    \includegraphics[width=6in]{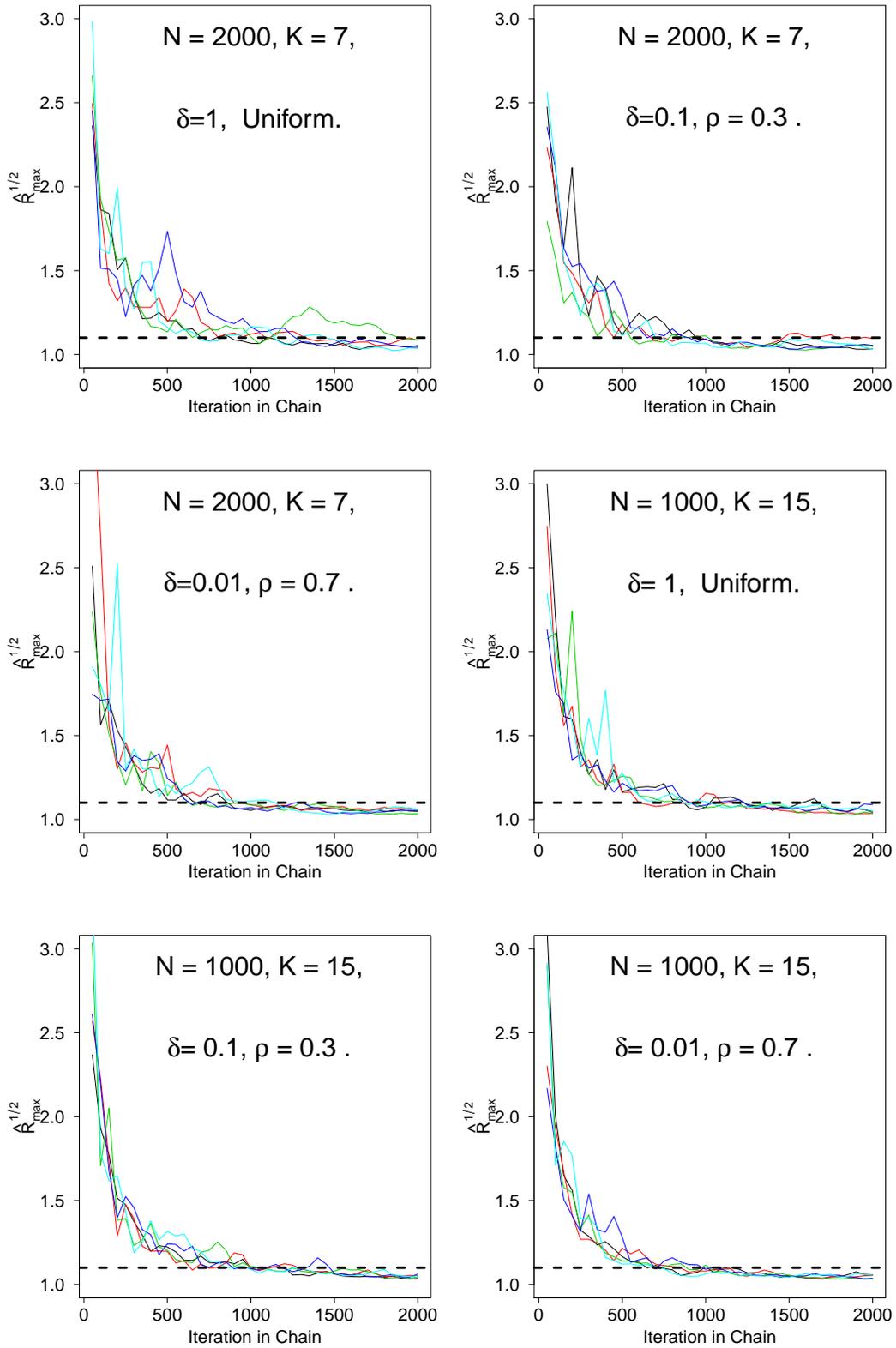}
    \caption{Plots the maximum of all potential scale reduction factors. The horizontal dashed line represents the value 1.1. }
    \label{Convergence}
\end{figure}

For the low dimension ($K=3,5$), the Dirichlet prior's hyper-parameter $\bm\delta$ is the $C$-dimensional vector $\bm 1$, leading to a non-informative prior. 
For the high dimension ($K=7, 15$),   three   $\bm\delta$'s are used: $\bm \delta= \bm{0.01},\bm{0.1}$ and $\bm{1}$, which are indicated as   “S”, “M” and “L” in Table  \ref{simulation settings}. 
For the GDINA model, the priors for the item parameter $\bm\lambda_j$ are shown as follows:
\begin{equation}
\lambda^{(w)}_j \sim
\begin{cases}\label{GDIP}
N(-1.2,0.4^2), \quad \ w=0,\\
N(0.9,0.3^2)/w,\quad w = 1,\cdots,K_j^*
\end{cases}
\end{equation}
which have a similar support set and a large variance compared with the generation regime for $w>0$. For the DINA model, non-informative Beta priors are taken for the slipping and guessing parameters. 
For fair comparison, all methods use the same set of priors. 

Besides the priors, we need to specify initial values for $\bm \alpha, \bm \pi$ and $\bm\lambda$. The initial value of $\bm\alpha$ is that each attribute is randomly sampled from an independent Bernoulli(0.5). The initial value of population parameter $\bm\pi$ is the $C$-dimensional vector $\bm {1/C}$. The initial value of $\bm\lambda$ is a random sample from the $\bm\lambda$'s prior. 

\cite{culpepper2015bayesian} showed that, for the DINA model, simultaneous Gibbs sampling only needed about 750 iterations to reach convergence. 
Consequently, in this paper, a 2000 iterations Markov chain is run and we discard the first 1000 iterations as burn-in which are adequate to reach convergence. For the GDINA model, we conduct a Markov chain with length 3000 and burn-in  length 2000. 

For the simulation results, potential scale reduction factor $\hat{R}^{1/2}$ \citep{gelman1992inference, brooks1998general} is used for convergence diagnosis. \cite{brooks1998general} suggested that $\hat{R}^{1/2}<1.2$ for all model parameters indicates that convergence has been reached. To make the conclusion more reliable, the condition  $\hat{R}^{1/2}<1.1$ can be used. In Figure \ref{Convergence}, the plots of the maximum of all potential scale reduction factors are shown for six representative conditions. The six conditions are from high-dimensional GDINA models. For each condition, we generate 5 independent response data. For each response data, 5 parallel chains are run to calculate the potential scale reduction factors for all item and population parameters. The plots are based upon the first 2000 chain lengths (from 50 to 2000 by increments of 50). The results in Figure \ref{Convergence} (and other diagnosis results not presented here) support that all chains can reach convergence   after 2000 iterations under the GDINA model.

\subsection{Simulation Results}
There are 7 tables to show the results for different settings of the DINA and GDINA models. For the GDINA model  with a probit link function, the generation of item parameters has a great probability that the intercept is negative and the interaction is positive. Consequently the results of the sampling with or without truncation are very similar, and here we only show the results of sampling with truncation. 
The digits of 
bias, RMSE, MSE, AAR and PAR$n$
are rounded off to four decimal places and the digits of time are rounded off to five significant figures. 

Tables \ref{LU} -- \ref{LC2} show the results of the uniform population, correlated structure with $\rho= 0.3$ and $0.7$ for the low dimension cases, respectively. Hereafter, the ``Sim", ``Seq", ``Ind", and ``sSim" represent the simultaneous (implemented by JAGS), sequential, independent, and self-compiled simultaneous Gibbs sampling methods, respectively. For the DINA model, `Sim", ``Seq" and ``sSim"  always obtain similar estimation results for item parameters, population parameters, and attribute patterns, which indicates the estimation consistency  between the sequential and simultaneous Gibbs sampling methods. When the population is uniform,  the ``Ind" method also performs similarly to the other methods. 
However, as the correlation $\rho$ among the attributes  increases, the ``Ind" method performs more poorly in  $\mathrm{\overline{MN}}\bm{_\pi}$.
For the GDINA model, inferences between the sequential and simultaneous Gibbs sampling methods are also consistent. 
As the correlation $\rho$ increases, the ``Ind" method will underperform the other methods, due to the violation of its assumption on the independence of the attributes.

For both models, the accuracy of estimations becomes better with the large sample size.  The classification accuracy of the ``Sim" method, as a baseline, is comparable to that of the other methods.

Tables \ref{HU} -- \ref{HC2} show the results of the uniform population and correlated structures with $\rho= 0.3$ and $0.7$ for the high dimension cases. 
 If the population distribution of attribute profiles is uniform, 
the ``Ind" method is comparable or slightly outperforms the sequential Gibbs sampling method. When the attributes become more correlated, 
the sequential sampling method outperforms the ``Ind" method.
For instance, in the high-dimensional condition with $\rho= 0.7$ and $K=15$, the sequential sampling with a smaller $\bm \delta$  is   superior to  the other methods.
We also find some common phenomena that when $K=15$, the $\mathrm{{\overline {RMSE}}\bm{_\pi}}$ is approximate to 0.0001 (i.e., the $\mathrm{{\overline {MSE}}\bm{_\pi}}$ is approximate to 0), which is due to the large number of latent classes. 


In Table \ref{Time}, the average computational time is reported. Since the dependent structure of the attribute profiles has almost negligible influence on the computational time, we only report the calculation time of different methods under the uniform population. The results show that the simultaneous Gibbs sampling implemented by JAGS has the slowest speed among the four methods. In order to show the computational superiority of the sequential Gibbs than that of the simultaneous Gibbs sampling, we also add the $K=7$ condition for the self-compiled simultaneous Gibbs sampling. Comparing the time of ``Seq" and ``sSim", when $K = 7$, the superiority of the sequential Gibbs sampling method appears in the GDINA model. However, the superiority of the sequential Gibbs sampling method has been reflected in the DINA model with $K=5$. In addition, ``Ind" is also   computationally efficient, with comparable computational time to the sequential Gibbs method.

 In the simulation study, using JAGS to estimate the GDINA model needs more memory space than the DINA model.
Particularly, for the GDINA model, 
we find that the estimation process in JAGS is often  executed   or   shut down due to “run out of memory”. 
Therefore, the presented  computational time for the GDINA model is only the average of those well-converged replications in JAGS, which leads to a counter-intuitive observation that  the simultaneous Gibbs sampling for the GDINA model is ``faster" than the simultaneous Gibbs sampling of the DINA model. 
Due to the high computational cost, the simultaneous Gibbs sampling time
for some high dimension cases is not reported.

Through simulation studies, we can find
when $K$ is small, the sequential Gibbs can use less time to obtain the results with similar accuracy as JAGS. When $K$ is large, the sequential Gibbs sampling algorithm still works well, but the simultaneous Gibbs sampling algorithm doesn't due to the high computational cost. The speed advantage of the sequential Gibbs sampling  become more apparent as $K$ increases.  In addition, when 
  the number of attributes $K$ goes large, the moderately small hyperparameter $\bm \delta$ is preferred. 
 Comparing  the independent and sequential Gibbs sampling methods, for the uniform structure with independent attributes, the independent Gibbs sampling method is comparable or slightly outperforms the sequential Gibbs sampling method. However, for correlated structures, the sequential Gibbs sampling method provides better performance  (see the Bias in Tables 5 and 8).

\begin{sidewaystable}[bhtp]

\renewcommand\tabcolsep{3.5pt} 
\small
\caption{Parameter recovery for Low dimension with the Uniform Population}
\begin{center}
\begin{threeparttable}
\begin{tabular}{llllllllllllllllllllll}
\hline
\multicolumn{2}{r}{ Attribute Number}&\multicolumn{8}{c}{$K=3$}&\multicolumn{8}{c}{$K=5$} \\ \cmidrule(lr){3-10}\cmidrule(lr){11-18}
\multicolumn{2}{r}{Sample Size}&\multicolumn{4}{c}{1000}&\multicolumn{4}{c}{2000}&\multicolumn{4}{c}{1000}&\multicolumn{4}{c}{2000}\\ \cmidrule(lr){3-6}\cmidrule(lr){7-10}\cmidrule(lr){11-14}\cmidrule(lr){15-18}
\multicolumn{2}{r}{Method}&Sim&Seq&Ind&sSim&Sim&Seq&Ind&sSim&Sim&Seq&Ind&sSim&Sim&Seq&Ind&sSim\\
\hline 
DINA&$\mathrm{\overline{Bias}}\bm{_g}$ & 0.0010 & 0.0010 & 0.0013 & 0.0010 & 0.0002 & 0.0002 & 0.0004 & 0.0002 & 0.0009 & 0.0011 & 0.0005 & 0.0011 & 0.0008 & 0.0009 & 0.0014 & 0.0009 \\
&$\mathrm{\overline{RMSE}}\bm{_g}$ & 0.0156 & 0.0156 & 0.0156 & 0.0155 & 0.0113 & 0.0113 & 0.0113 & 0.0114 & 0.0164 & 0.0164 & 0.0161 & 0.0164 & 0.0122 & 0.0122 & 0.0122 & 0.0122 \\ 
&$\mathrm{\overline{MSE}}\bm{_g}$ & 0.0003 & 0.0003 & 0.0003 & 0.0003 & 0.0001 & 0.0001 & 0.0001 & 0.0001 & 0.0003 & 0.0003 & 0.0003 & 0.0003 & 0.0002 & 0.0002 & 0.0002 & 0.0002 \\
&$\mathrm{\overline{Bias}}\bm{_s}$ & 0.0046 & 0.0037 & 0.0034 & 0.0037 & 0.0001 & -0.0003 & -0.0004 & -0.0003 & 0.0035 & 0.0022 & 0.0036 & 0.0021 & 0.0019 & 0.0013 & 0.0005 & 0.0013 \\
&$\mathrm{\overline{RMSE}}\bm{_s}$ & 0.0257 & 0.0254 & 0.0253 & 0.0254 & 0.0171 & 0.0171 & 0.0171 & 0.0171 & 0.0278 & 0.0273 & 0.0273 & 0.0274 & 0.0179 & 0.0178 & 0.0177 & 0.0178 \\
&$\mathrm{\overline{MSE}}\bm{_s}$ & 0.0007 & 0.0007 & 0.0007 & 0.0007 & 0.0003 & 0.0003 & 0.0003 & 0.0003 & 0.0009 & 0.0008 & 0.0008 & 0.0008 & 0.0004 & 0.0004 & 0.0003 & 0.0004 \\ 
&$\mathrm{\overline{MN}}\bm{_\pi}$ & 0.0158 & 0.0159 & 0.0144 & 0.0159 & 0.0117 & 0.0116 & 0.0115 & 0.0118 & 0.0154 & 0.0155 & 0.0109 & 0.0154 & 0.0105 & 0.0105 & 0.0065 & 0.0106 \\
&$\mathrm{\overline{RMSE}}\bm{_\pi}$ & 0.0094 & 0.0093 & 0.0082 & 0.0094 & 0.0060 & 0.0059 & 0.0054 & 0.0060 & 0.0059 & 0.0059 & 0.0041 & 0.0059 & 0.0040 & 0.0040 & 0.0026 & 0.0040 \\
&$\mathrm{\overline{MSE}}\bm{_\pi}$ & 0.0001 & 0.0001 & 0.0001 & 0.0001 & 0.0000 & 0.0000 & 0.0000 & 0.0000 & 0.0000 & 0.0000 & 0.0000 & 0.0000 & 0.0000 & 0.0000 & 0.0000 & 0.0000 \\
&AAR & 0.9742 & 0.9741 & 0.9739 & 0.9740 & 0.9749 & 0.9748 & 0.9748 & 0.9748 & 0.9445 & 0.9443 & 0.9445 & 0.9443 & 0.9461 & 0.9459 & 0.9466 & 0.9460 \\ 
&PAR1 & 0.9956 & 0.9953 & 0.9952 & 0.9954 & 0.9962 & 0.9963 & 0.9961 & 0.9961 & 0.9644 & 0.9643 & 0.9654 & 0.9645 & 0.9667 & 0.9669 & 0.9676 & 0.9667 \\
GDINA&$\mathrm{\overline{Bias}}\bm{_\lambda}$& 0.0305 & 0.0310 & 0.0297 & 0.0307 & 0.0243 & 0.0244 & 0.0250 & 0.0244 & 0.0371 & 0.0372 & 0.0313 & 0.0374 & 0.0326 & 0.0331 & 0.0319 & 0.0332 \\ 
&$\mathrm{\overline{RMSE}}\bm{_\lambda}$ & 0.1370 & 0.1373 & 0.1378 & 0.1370 & 0.1148 & 0.1149 & 0.1151 & 0.1148 & 0.1334 & 0.1336 & 0.1325 & 0.1336 & 0.1288 & 0.1288 & 0.1275 & 0.1285 \\ 
&$\mathrm{\overline{MSE}}\bm{_\lambda}$ & 0.0221 & 0.0222 & 0.0223 & 0.0221 & 0.0162 & 0.0161 & 0.0162 & 0.0161 & 0.0205 & 0.0205 & 0.0203 & 0.0206 & 0.0209 & 0.0210 & 0.0206 & 0.0209 \\ 
&$\mathrm{\overline{MN}}\bm{_\pi}$  & 0.0235 & 0.0238 & 0.0180 & 0.0236 & 0.0119 & 0.0118 & 0.0091 & 0.0119 & 0.0214 & 0.0217 & 0.0107 & 0.0214 & 0.0153 & 0.0152 & 0.0063 & 0.0155 \\ 
&$\mathrm{\overline{RMSE}}\bm{_\pi}$ & 0.0129 & 0.0130 & 0.0098 & 0.0130 & 0.0063 & 0.0063 & 0.0046 & 0.0063 & 0.0089 & 0.0090 & 0.0040 & 0.0090 & 0.0061 & 0.0061 & 0.0025 & 0.0061 \\ 
&$\mathrm{\overline{MSE}}\bm{_\pi}$  & 0.0002 & 0.0002 & 0.0001 & 0.0002 & 0.0000 & 0.0000 & 0.0000 & 0.0000 & 0.0001 & 0.0001 & 0.0000 & 0.0001 & 0.0000 & 0.0000 & 0.0000 & 0.0000 \\ 
&AAR & 0.9436 & 0.9423 & 0.9431 & 0.9435 & 0.9452 & 0.9446 & 0.9443 & 0.9452 & 0.8921 & 0.8920 & 0.8945 & 0.8923 & 0.8959 & 0.8960 & 0.8969 & 0.8956 \\ 
&PAR1 & 0.9575 & 0.9557 & 0.9563 & 0.9578 & 0.9604 & 0.9595 & 0.9590 & 0.9605 & 0.8882 & 0.8876 & 0.8909 & 0.8886 & 0.8891 & 0.8892 & 0.8907 & 0.8887 \\ 

\hline
\end{tabular}

\begin{tablenotes}
        \footnotesize
        \item \textbf{Note}. The ``Sim", ``Seq", ``Ind" and ``sSim" represent the simultaneous (implemented by JAGS), sequential, independent and self-compiled simultaneous Gibbs sampling methods, respectively.
              \end{tablenotes}
      \end{threeparttable}
\end{center}
\label{LU}
\end{sidewaystable}%

\begin{sidewaystable}[htp]
\renewcommand\tabcolsep{3.5pt} 
\small
\caption{Parameter recovery for Low dimension with the Correlation $\rho=0.3$}
\begin{center}
\begin{threeparttable}
\begin{tabular}{llllllllllllllllllllll}
\hline
\multicolumn{2}{r}{ Attribute Number}&\multicolumn{8}{c}{$K=3$}&\multicolumn{8}{c}{$K=5$} \\ \cmidrule(lr){3-10}\cmidrule(lr){11-18}
\multicolumn{2}{r}{Sample Size}&\multicolumn{4}{c}{1000}&\multicolumn{4}{c}{2000}&\multicolumn{4}{c}{1000}&\multicolumn{4}{c}{2000}\\ \cmidrule(lr){3-6}\cmidrule(lr){7-10}\cmidrule(lr){11-14}\cmidrule(lr){15-18}
\multicolumn{2}{r}{Method}&Sim&Seq&Ind&sSim&Sim&Seq&Ind&sSim&Sim&Seq&Ind&sSim&Sim&Seq&Ind&sSim\\
\hline
DINA &$\mathrm{\overline{Bias}}\bm{_g}$ & 0.0004 & 0.0005 & -0.0018 & 0.0005 & -0.0002 & -0.0001 & -0.0024 & -0.0001 & 0.0010 & 0.0012 & -0.0044 & 0.0011 & 0.0005 & 0.0006 & -0.0043 & 0.0006 \\
&$\mathrm{\overline{RMSE}}\bm{_g}$ & 0.0159 & 0.0159 & 0.0161 & 0.0159 & 0.0113 & 0.0113 & 0.0117 & 0.0113 & 0.0170 & 0.0170 & 0.0180 & 0.0170 & 0.0120 & 0.0120 & 0.0132 & 0.0120 \\
&$\mathrm{\overline{MSE}}\bm{_g}$ & 0.0003 & 0.0003 & 0.0003 & 0.0003 & 0.0001 & 0.0001 & 0.0001 & 0.0001 & 0.0003 & 0.0003 & 0.0003 & 0.0003 & 0.0002 & 0.0002 & 0.0002 & 0.0001 \\
&$\mathrm{\overline{Bias}}\bm{_s}$ & 0.0023 & 0.0016 & 0.0037 & 0.0016 & 0.0006 & 0.0003 & 0.0024 & 0.0003 & 0.0030 & 0.0020 & 0.0072 & 0.0020 & 0.0012 & 0.0007 & 0.0046 & 0.0007 \\
&$\mathrm{\overline{RMSE}}\bm{_s}$ & 0.0216 & 0.0215 & 0.0217 & 0.0215 & 0.0157 & 0.0156 & 0.0161 & 0.0157 & 0.0237 & 0.0235 & 0.0250 & 0.0235 & 0.0166 & 0.0166 & 0.0179 & 0.0166 \\ 
&$\mathrm{\overline{MSE}}\bm{_s}$ & 0.0005 & 0.0005 & 0.0005 & 0.0005 & 0.0003 & 0.0003 & 0.0003 & 0.0003 & 0.0006 & 0.0006 & 0.0007 & 0.0006 & 0.0003 & 0.0003 & 0.0003 & 0.0003 \\
&$\mathrm{\overline{MN}}\bm{_\pi}$ & 0.0076 & 0.0077 & 0.0227 & 0.0076 & 0.0055 & 0.0054 & 0.0253 & 0.0054 & 0.0128 & 0.0128 & 0.0480 & 0.0125 & 0.0084 & 0.0082 & 0.0444 & 0.0083 \\ 
&$\mathrm{\overline{RMSE}}\bm{_\pi}$ & 0.0034 & 0.0034 & 0.0068 & 0.0034 & 0.0025 & 0.0025 & 0.0072 & 0.0025 & 0.0039 & 0.0039 & 0.0047 & 0.0039 & 0.0026 & 0.0026 & 0.0039 & 0.0026 \\
&$\mathrm{\overline{MSE}}\bm{_\pi}$  & 0.0000 & 0.0000 & 0.0001 & 0.0000 & 0.0000 & 0.0000 & 0.0001 & 0.0000 & 0.0000 & 0.0000 & 0.0001 & 0.0000 & 0.0000 & 0.0000 & 0.0001 & 0.0000 \\
& AAR & 0.9789 & 0.9788 & 0.9748 & 0.9789 & 0.9790 & 0.9789 & 0.9740 & 0.9790 & 0.9531 & 0.9530 & 0.9423 & 0.9533 & 0.9545 & 0.9543 & 0.9445 & 0.9544 \\ 
& APR1 & 0.9964 & 0.9965 & 0.9964 & 0.9964 & 0.9969 & 0.9969 & 0.9967 & 0.9969 & 0.9724 & 0.9722 & 0.9624 & 0.9721 & 0.9732 & 0.9728 & 0.9639 & 0.9728 \\ 
GDINA & $\mathrm{\overline{Bias}}\bm{_\lambda}$ & 0.0304 & 0.0314 & 0.0371 & 0.0310 & 0.0220 & 0.0224 & 0.0279 & 0.0225 & 0.0371 & 0.0398 & 0.0565 & 0.0402 & 0.0270 & 0.0285 & 0.0457 & 0.0281 \\ 
& $\mathrm{\overline{RMSE}}\bm{_\lambda}$  & 0.1301 & 0.1303 & 0.1285 & 0.1304 & 0.1051 & 0.1053 & 0.1021 & 0.1052 & 0.1457 & 0.1459 & 0.1474 & 0.1461 & 0.1192 & 0.1193 & 0.1200 & 0.1194 \\ 
& $\mathrm{\overline{MSE}}\bm{_\lambda}$  & 0.0198 & 0.0199 & 0.0195 & 0.0199 & 0.0130 & 0.0130 & 0.0126 & 0.0130 & 0.0250 & 0.0252 & 0.0260 & 0.0252 & 0.0176 & 0.0176 & 0.0183 & 0.0176 \\ 
& $\mathrm{\overline{MN}}\bm{_\pi}$  & 0.0109 & 0.0111 & 0.0250 & 0.0109 & 0.0091 & 0.0093 & 0.0195 & 0.0091 & 0.0215 & 0.0218 & 0.0453 & 0.0223 & 0.0153 & 0.0157 & 0.0415 & 0.0156 \\ 
& $\mathrm{\overline{RMSE}}\bm{_\pi}$ & 0.0064 & 0.0065 & 0.0112 & 0.0063 & 0.0049 & 0.0050 & 0.0089 & 0.0049 & 0.0076 & 0.0076 & 0.0073 & 0.0077 & 0.0056 & 0.0057 & 0.0064 & 0.0057 \\ 
& $\mathrm{\overline{MSE}}\bm{_\pi}$ & 0.0000 & 0.0000 & 0.0002 & 0.0000 & 0.0000 & 0.0000 & 0.0001 & 0.0000 & 0.0001 & 0.0001 & 0.0001 & 0.0001 & 0.0000 & 0.0000 & 0.0001 & 0.0000 \\ 
&AAR & 0.9488 & 0.9485 & 0.9466 & 0.9488 & 0.9531 & 0.9531 & 0.9512 & 0.9531 & 0.8936 & 0.8933 & 0.8914 & 0.8933 & 0.8976 & 0.8977 & 0.8938 & 0.8977 \\ 
&PAR1 & 0.9660 & 0.9657 & 0.9630 & 0.9661 & 0.9701 & 0.9697 & 0.9665 & 0.9697 & 0.8931 & 0.8920 & 0.8877 & 0.8924 & 0.9031 & 0.9030 & 0.8952 & 0.9030 \\ 
\hline
\end{tabular}
\begin{tablenotes}
        \footnotesize
               \item \textbf{Note}. The ``Sim", ``Seq", ``Ind" and ``sSim" represent the simultaneous (implemented by JAGS), sequential, independent and self-compiled simultaneous Gibbs sampling methods, respectively. 
      \end{tablenotes}
      \end{threeparttable}
\end{center}
\label{LC1}
\end{sidewaystable}%

 \begin{sidewaystable}[htp]
 \renewcommand\tabcolsep{3.5pt} 
\small
\caption{Parameter recovery for Low dimension with the Correlation $\rho=0.7$}
\begin{center}
\begin{threeparttable}
\begin{tabular}{llllllllllllllllllllll}
\hline
\multicolumn{2}{r}{ Attribute Number}&\multicolumn{8}{c}{$K=3$}&\multicolumn{8}{c}{$K=5$} \\ \cmidrule(lr){3-10}\cmidrule(lr){11-18}
\multicolumn{2}{r}{Sample Size}&\multicolumn{4}{c}{1000}&\multicolumn{4}{c}{2000}&\multicolumn{4}{c}{1000}&\multicolumn{4}{c}{2000}\\ \cmidrule(lr){3-6}\cmidrule(lr){7-10}\cmidrule(lr){11-14}\cmidrule(lr){15-18}
\multicolumn{2}{r}{Method}&Sim&Seq&Ind&sSim&Sim&Seq&Ind&sSim&Sim&Seq&Ind&sSim&Sim&Seq&Ind&sSim\\
\hline
DINA &$\mathrm{\overline{Bias}}\bm{_g}$ & 0.0001 & 0.0001 & -0.0055 & 0.0002 & 0.0000 & 0.0000 & -0.0057 & 0.0000 & 0.0019 & 0.0019 & -0.0097 & 0.0019 & -0.0000 & -0.0000 & -0.0119 & -0.0000 \\ 
&$\mathrm{\overline{RMSE}}\bm{_g}$ & 0.0164 & 0.0164 & 0.0181 & 0.0164 & 0.0116 & 0.0116 & 0.0140 & 0.0116 & 0.0169 & 0.0169 & 0.0223 & 0.0169 & 0.0130 & 0.0130 & 0.0200 & 0.0130 \\ 
&$\mathrm{\overline{MSE}}\bm{_g}$ & 0.0003 & 0.0003 & 0.0003 & 0.0003 & 0.0001 & 0.0001 & 0.0002 & 0.0001 & 0.0003 & 0.0003 & 0.0005 & 0.0003 & 0.0002 & 0.0002 & 0.0005 & 0.0002 \\ 
&$\mathrm{\overline{Bias}}\bm{_s}$ & 0.0022 & 0.0016 & 0.0069 & 0.0016 & 0.0011 & 0.0008 & 0.0066 & 0.0008 & 0.0019 & 0.0014 & 0.0108 & 0.0014 & 0.0008 & 0.0005 & 0.0104 & 0.0005 \\ 
&$\mathrm{\overline{RMSE}}\bm{_s}$ & 0.0209 & 0.0208 & 0.0227 & 0.0208 & 0.0145 & 0.0145 & 0.0171 & 0.0145 & 0.0210 & 0.0209 & 0.0260 & 0.0209 & 0.0141 & 0.0140 & 0.0210 & 0.0140 \\
&$\mathrm{\overline{MSE}}\bm{_s}$ & 0.0004 & 0.0004 & 0.0005 & 0.0004 & 0.0002 & 0.0002 & 0.0003 & 0.0002 & 0.0005 & 0.0004 & 0.0007 & 0.0004 & 0.0002 & 0.0002 & 0.0005 & 0.0002 \\ 
&$\mathrm{\overline{MN}}\bm{_\pi}$ & 0.0072 & 0.0070 & 0.0546 & 0.0070 & 0.0054 & 0.0053 & 0.0581 & 0.0053 & 0.0127 & 0.0128 & 0.1206 & 0.0126 & 0.0080 & 0.0082 & 0.1259 & 0.0082 \\ 
&$\mathrm{\overline{RMSE}}\bm{_\pi}$ & 0.0034 & 0.0035 & 0.0144 & 0.0034 & 0.0024 & 0.0024 & 0.0149 & 0.0024 & 0.0033 & 0.0033 & 0.0087 & 0.0033 & 0.0023 & 0.0023 & 0.0087 & 0.0023 \\ 
&$\mathrm{\overline{MSE}}\bm{_\pi}$ & 0.0000 & 0.0000 & 0.0005 & 0.0000 & 0.0000 & 0.0000 & 0.0005 & 0.0000 & 0.0000 & 0.0000 & 0.0005 & 0.0000 & 0.0000 & 0.0000 & 0.0005 & 0.0000 \\ 
&AAR & 0.9815 & 0.9817 & 0.9696 & 0.9817 & 0.9825 & 0.9825 & 0.9678 & 0.9826 & 0.9652 & 0.9651 & 0.9380 & 0.9651 & 0.9667 & 0.9667 & 0.9361 & 0.9667 \\ 
&PAR1 & 0.9966 & 0.9968 & 0.9959 & 0.9967 & 0.9976 & 0.9976 & 0.9963 & 0.9975 & 0.9795 & 0.9796 & 0.9511 & 0.9796 & 0.9819 & 0.9821 & 0.9478 & 0.9819 \\
GDINA&$\mathrm{\overline{Bias}}\bm{_\lambda}$ & 0.0247 & 0.0252 & 0.0343 & 0.0252 & 0.0177 & 0.0186 & 0.0308 & 0.0186 & 0.0197 & 0.0190 & 0.0619 & 0.0184 & 0.0138 & 0.0141 & 0.0477 & 0.0141 \\ 
&$\mathrm{\overline{RMSE}}\bm{_\lambda}$  & 0.1227 & 0.1226 & 0.1347 & 0.1228 & 0.1139 & 0.1142 & 0.1356 & 0.1142 & 0.1458 & 0.1462 & 0.1647 & 0.1466 & 0.1145 & 0.1140 & 0.1406 & 0.1141 \\ 
&$\mathrm{\overline{MSE}}\bm{_\lambda}$  & 0.0173 & 0.0173 & 0.0209 & 0.0173 & 0.0164 & 0.0164 & 0.0232 & 0.0164 & 0.0262 & 0.0262 & 0.0339 & 0.0264 & 0.0166 & 0.0164 & 0.0259 & 0.0165 \\ 
&$\mathrm{\overline{MN}}\bm{_\pi}$  & 0.0117 & 0.0117 & 0.0754 & 0.0118 & 0.0076 & 0.0075 & 0.0762 & 0.0077 & 0.0198 & 0.0205 & 0.1253 & 0.0195 & 0.0129 & 0.0132 & 0.1414 & 0.0125 \\ 
&$\mathrm{\overline{RMSE}}\bm{_\pi}$  & 0.0069 & 0.0069 & 0.0269 & 0.0069 & 0.0042 & 0.0043 & 0.0285 & 0.0043 & 0.0057 & 0.0058 & 0.0153 & 0.0057 & 0.0041 & 0.0041 & 0.0149 & 0.0041 \\ 
&$\mathrm{\overline{MSE}}\bm{_\pi}$  & 0.0000 & 0.0000 & 0.0011 & 0.0000 & 0.0000 & 0.0000 & 0.0012 & 0.0000 & 0.0000 & 0.0000 & 0.0009 & 0.0000 & 0.0000 & 0.0000 & 0.0010 & 0.0000 \\ 
&AAR & 0.9561 & 0.9562 & 0.9449 & 0.9559 & 0.9629 & 0.9626 & 0.9462 & 0.9628 & 0.9365 & 0.9362 & 0.9150 & 0.9364 & 0.9368 & 0.9369 & 0.9137 & 0.9369 \\ 
&PAR1 & 0.9774 & 0.9780 & 0.9720 & 0.9776 & 0.9790 & 0.9786 & 0.9724 & 0.9789 & 0.9481 & 0.9475 & 0.9286 & 0.9480 & 0.9514 & 0.9516 & 0.9321 & 0.9514 \\ 
\hline
\end{tabular}
\begin{tablenotes}
        \footnotesize
        \item \textbf{Note}. The ``Sim", ``Seq", ``Ind" and ``sSim" represent the simultaneous (implemented by JAGS), sequential, independent and self-compiled simultaneous Gibbs sampling methods, respectively.
      \end{tablenotes}
      \end{threeparttable}
\end{center}
\label{LC2}
\end{sidewaystable}%

\begin{sidewaystable}[htp]
\renewcommand\tabcolsep{3.5pt} 
\small
\caption{Parameter recovery for High dimension with the Uniform Population}
\begin{center}
\begin{spacing}{1}
\begin{threeparttable}
\begin{tabular}{llllllllllllllllll}
\hline
\multicolumn{2}{r}{Attribute Number}&\multicolumn{8}{c}{$K=7$}&\multicolumn{8}{c}{$K=15$} \\ \cmidrule(lr){3-10}\cmidrule(lr){11-18}
\multicolumn{2}{r}{Sample Size}&\multicolumn{4}{c}{1000}&\multicolumn{4}{c}{2000}&\multicolumn{4}{c}{1000}&\multicolumn{4}{c}{2000}\\ \cmidrule(lr){3-6}\cmidrule(lr){7-10}\cmidrule(lr){11-14}\cmidrule(lr){15-18}
\multicolumn{2}{r}{Method}&$\bm{0.01}$&$\bm {0.1}$&$\bm {1}$&Ind&$\bm{0.01}$&$\bm {0.1}$&$\bm {1}$&Ind&$\bm{0.01}$&$\bm {0.1}$&$\bm{1}$&Ind&$\bm{0.01}$&$\bm{0.1}$&$\bm{1}$&Ind\\
\hline

DINA &$\mathrm{\overline{Bias}}\bm{_g}$ & 0.0066 & 0.0041 & 0.0026 & 0.0018 & 0.0018 & 0.0010 & 0.0005 & -0.0004 & 0.0127 & 0.0055 & 0.0056 & 0.0055 & 0.0120 & 0.0007 & 0.0001 & 0.0001 \\ 
&$\mathrm{\overline{RMSE}}\bm{_g}$ & 0.0208 & 0.0190 & 0.0182 & 0.0178 & 0.0140 & 0.0134 & 0.0131 & 0.0128 & 0.0329 & 0.0299 & 0.0290 & 0.0293 & 0.0260 & 0.0205 & 0.0200 & 0.0198 \\ 
&$\mathrm{\overline{MSE}}\bm{_g}$ & 0.0005 & 0.0004 & 0.0003 & 0.0003 & 0.0002 & 0.0002 & 0.0002 & 0.0002 & 0.0012 & 0.0010 & 0.0009 & 0.0010 & 0.0007 & 0.0005 & 0.0004 & 0.0004 \\
&$\mathrm{\overline{Bias}}\bm{_s}$ & 0.0112 & 0.0060 & 0.0026 & 0.0013 & 0.0073 & 0.0040 & 0.0020 & 0.0019 & 0.0036 & -0.0038 & -0.0046 & -0.0046 & 0.0079 & 0.0013 & 0.0013 & 0.0012 \\
&$\mathrm{\overline{RMSE}}\bm{_s}$ & 0.0321 & 0.0290 & 0.0280 & 0.0276 & 0.0229 & 0.0214 & 0.0208 & 0.0201 & 0.0394 & 0.0367 & 0.0359 & 0.0360 & 0.0321 & 0.0268 & 0.0262 & 0.0259 \\ 
&$\mathrm{\overline{MSE}}\bm{_s}$ & 0.0011 & 0.0009 & 0.0008 & 0.0008 & 0.0006 & 0.0005 & 0.0005 & 0.0005 & 0.0017 & 0.0015 & 0.0014 & 0.0014 & 0.0011 & 0.0008 & 0.0008 & 0.0007 \\ 
&$\mathrm{\overline{MN}}\bm{_\pi}$ & 0.0232 & 0.0163 & 0.0106 & 0.0056 & 0.0154 & 0.0114 & 0.0082 & 0.0040 & 0.0056 & 0.0003 & 0.0000 & 0.0006 & 0.0052 & 0.0004 & 0.0000 & 0.0003 \\
&$\mathrm{\overline{RMSE}}\bm{_\pi}$ & 0.0072 & 0.0052 & 0.0032 & 0.0018 & 0.0052 & 0.0039 & 0.0028 & 0.0013 & 0.0001 & 0.0000 & 0.0000 & 0.0000 & 0.0001 & 0.0000 & 0.0000 & 0.0000 \\ 
&$\mathrm{\overline{MSE}}\bm{_\pi}$ & 0.0001 & 0.0000 & 0.0000 & 0.0000 & 0.0000 & 0.0000 & 0.0000 & 0.0000 & 0.0000 & 0.0000 & 0.0000 & 0.0000 & 0.0000 & 0.0000 & 0.0000 & 0.0000 \\ 
&AAR & 0.9039 & 0.9178 & 0.9256 & 0.9173 & 0.9161 & 0.9237 & 0.9269 & 0.9200 & 0.8167 & 0.8386 & 0.8396 & 0.8246 & 0.8114 & 0.8388 & 0.8402 & 0.8201 \\ 
&PAR2 & 0.9712 & 0.9780 & 0.9792 & 0.9777 & 0.9773 & 0.9794 & 0.9799 & 0.9779 & 0.4679 & 0.5574 & 0.5632 & 0.4998 & 0.4436 & 0.5585 & 0.5654 & 0.4797 \\
GDINA&$\mathrm{\overline{Bias}}\bm{_\lambda}$ & 0.0372 & 0.0410 & 0.0403 & 0.0355 & 0.0197 & 0.0264 & 0.0304 & 0.0320 & 0.0336 & 0.0325 & 0.0325 & 0.0320 & 0.0245 & 0.0315 & 0.0318 & 0.0324 \\ 
&$\mathrm{\overline{RMSE}}\bm{_\lambda}$  & 0.1884 & 0.1700 & 0.1595 & 0.1570 & 0.1445 & 0.1328 & 0.1256 & 0.1219 & 0.2057 & 0.1804 & 0.1781 & 0.1784 & 0.1886 & 0.1578 & 0.1562 & 0.1552 \\ 
&$\mathrm{\overline{MSE}}\bm{_\lambda}$  & 0.0424 & 0.0344 & 0.0306 & 0.0296 & 0.0239 & 0.0204 & 0.0184 & 0.0175 & 0.0465 & 0.0367 & 0.0359 & 0.0361 & 0.0411 & 0.0297 & 0.0293 & 0.0290 \\ 
&$\mathrm{\overline{MN}}\bm{_\pi}$  & 0.0473 & 0.0311 & 0.0144 & 0.0037 & 0.0253 & 0.0185 & 0.0107 & 0.0025 & 0.0077 & 0.0000 & 0.0000 & 0.0001 & 0.0090 & 0.0001 & 0.0000 & 0.0000 \\ 
&$\mathrm{\overline{RMSE}}\bm{_\pi}$  & 0.0120 & 0.0077 & 0.0036 & 0.0012 & 0.0089 & 0.0060 & 0.0033 & 0.0009 & 0.0001 & 0.0000 & 0.0000 & 0.0000 & 0.0001 & 0.0000 & 0.0000 & 0.0000 \\ 
&$\mathrm{\overline{MSE}}\bm{_\pi}$  & 0.0002 & 0.0001 & 0.0000 & 0.0000 & 0.0001 & 0.0000 & 0.0000 & 0.0000 & 0.0000 & 0.0000 & 0.0000 & 0.0000 & 0.0000 & 0.0000 & 0.0000 & 0.0000 \\ 
& AAR & 0.7846 & 0.8187 & 0.8358 & 0.8401 & 0.8298 & 0.8504 & 0.8589 & 0.8625 & 0.6852 & 0.7328 & 0.7355 & 0.7356 & 0.6754 & 0.7266 & 0.7303 & 0.7302 \\ 
&PAR2   & 0.8187 & 0.8760 & 0.8943 & 0.9014 & 0.8981 & 0.9212 & 0.9278 & 0.9305 & 0.1025 & 0.1851 & 0.1923 & 0.1950 & 0.0897 & 0.1781 & 0.1880 & 0.1865 \\ 

\hline
\end{tabular}
\begin{tablenotes}
        \footnotesize
        \item \textbf{Note}. In the row “Method”, there are three different levels of $\bm \delta$; the sequential Gibbs samplings with $\bm \delta=\bm{0.01},\bm{0.1}$ and $\bm{1}$. 
      \end{tablenotes}
      \end{threeparttable}
      \end{spacing}
\end{center}
\label{HU}
\end{sidewaystable}%
\begin{sidewaystable}[htp]
\renewcommand\tabcolsep{3.5pt} 
\small
\caption{Parameter recovery for High Dimension with the Correlation $\rho=0.3$}
\begin{center}
\begin{threeparttable}
\begin{tabular}{llllllllllllllllll}
\hline
\multicolumn{2}{r}{Attribute Number}&\multicolumn{8}{c}{$K=7$}&\multicolumn{8}{c}{$K=15$} \\ \cmidrule(lr){3-10}\cmidrule(lr){11-18}
\multicolumn{2}{r}{Sample Size}&\multicolumn{4}{c}{1000}&\multicolumn{4}{c}{2000}&\multicolumn{4}{c}{1000}&\multicolumn{4}{c}{2000}\\ \cmidrule(lr){3-6}\cmidrule(lr){7-10}\cmidrule(lr){11-14}\cmidrule(lr){15-18}
\multicolumn{2}{r}{Method}&$\bm{0.01}$&$\bm {0.1}$&$\bm {1}$&Ind&$\bm{0.01}$&$\bm {0.1}$&$\bm {1}$&Ind&$\bm{0.01}$&$\bm {0.1}$&$\bm{1}$&Ind&$\bm{0.01}$&$\bm{0.1}$&$\bm{1}$&Ind\\
\hline

DINA &$\mathrm{\overline{Bias}}\bm{_g}$ & 0.0069 & 0.0039 & 0.0005 & -0.0046 & 0.0047 & 0.0029 & 0.0012 & -0.0038 & 0.0074 & -0.0053 & -0.0013 & 0.0008 & 0.0151 & -0.0011 & 0.0012 & 0.0049 \\ 
&$\mathrm{\overline{RMSE}}\bm{_g}$ &  0.0212 & 0.0193 & 0.0185 & 0.0200 & 0.0152 & 0.0142 & 0.0137 & 0.0149 & 0.0283 & 0.0290 & 0.0292 & 0.0293 & 0.0251 & 0.0189 & 0.0208 & 0.0212 \\ 
&$\mathrm{\overline{MSE}}\bm{_g}$ & 0.0005 & 0.0004 & 0.0004 & 0.0004 & 0.0003 & 0.0002 & 0.0002 & 0.0002 & 0.0008 & 0.0009 & 0.0010 & 0.0009 & 0.0007 & 0.0004 & 0.0005 & 0.0005 \\
&$\mathrm{\overline{Bias}}\bm{_s}$ & 0.0092 & 0.0058 & 0.0045 & 0.0053 & 0.0044 & 0.0025 & 0.0015 & 0.0022 & 0.0053 & 0.0000 & -0.0085 & -0.0111 & 0.0012 & -0.0034 & -0.0116 & -0.0163 \\ 
&$\mathrm{\overline{RMSE}}\bm{_s}$ & 0.0270 & 0.0251 & 0.0246 & 0.0265 & 0.0181 & 0.0173 & 0.0170 & 0.0194 & 0.0339 & 0.0335 & 0.0386 & 0.0399 & 0.0241 & 0.0227 & 0.0294 & 0.0326 \\ 
&$\mathrm{\overline{MSE}}\bm{_s}$ & 0.0008 & 0.0007 & 0.0006 & 0.0007 & 0.0003 & 0.0003 & 0.0003 & 0.0004 & 0.0012 & 0.0012 & 0.0018 & 0.0020 & 0.0006 & 0.0006 & 0.0012 & 0.0017 \\ 
&$\mathrm{\overline{MN}}\bm{_\pi}$ & 0.0233 & 0.0160 & 0.0110 & 0.0384 & 0.0174 & 0.0124 & 0.0096 & 0.0416 & 0.0177 & 0.0100 & 0.0128 & 0.0115 & 0.0174 & 0.0110 & 0.0188 & 0.0183 \\ 
&$\mathrm{\overline{RMSE}}\bm{_\pi}$ & 0.0055 & 0.0038 & 0.0025 & 0.0022 & 0.0041 & 0.0029 & 0.0020 & 0.0019 & 0.0001 & 0.0001 & 0.0001 & 0.0001 & 0.0001 & 0.0001 & 0.0001 & 0.0001 \\ 
&$\mathrm{\overline{MSE}}\bm{_\pi}$ & 0.0000 & 0.0000 & 0.0000 & 0.0000 & 0.0000 & 0.0000 & 0.0000 & 0.0000 & 0.0000 & 0.0000 & 0.0000 & 0.0000 & 0.0000 & 0.0000 & 0.0000 & 0.0000 \\
&AAR & 0.9114 & 0.9219 & 0.9287 & 0.9117 & 0.9199 & 0.9265 & 0.9306 & 0.9176 & 0.8347 & 0.8506 & 0.8448 & 0.8193 & 0.8314 & 0.8574 & 0.8499 & 0.8231 \\ 
&PAR2 & 0.9749 & 0.9790 & 0.9820 & 0.9701 & 0.9793 & 0.9818 & 0.9826 & 0.9734 & 0.5423 & 0.6028 & 0.5769 & 0.5039 & 0.5250 & 0.6302 & 0.6004 & 0.5161 \\
GDINA&$\mathrm{\overline{Bias}}\bm{_\lambda}$ & 0.0302 & 0.0289 & 0.0359 & 0.0459 & 0.0121 & 0.0200 & 0.0286 & 0.0508 & 0.0454 & 0.0492 & 0.0488 & 0.0490 & 0.0440 & 0.0585 & 0.0610 & 0.0602 \\ 
&$\mathrm{\overline{RMSE}}\bm{_\lambda}$ & 0.1867 & 0.1685 & 0.1581 & 0.1565 & 0.1469 & 0.1328 & 0.1235 & 0.1246 & 0.1940 & 0.1799 & 0.1788 & 0.1791 & 0.1730 & 0.1637 & 0.1693 & 0.1683 \\ 
&$\mathrm{\overline{MSE}}\bm{_\lambda}$ & 0.0412 & 0.0341 & 0.0300 & 0.0297 & 0.0256 & 0.0211 & 0.0186 & 0.0198 & 0.0429 & 0.0381 & 0.0381 & 0.0383 & 0.0342 & 0.0325 & 0.0352 & 0.0347 \\ 
&$\mathrm{\overline{MN}}\bm{_\pi}$ & 0.0510 & 0.0313 & 0.0207 & 0.0450 & 0.0351 & 0.0229 & 0.0172 & 0.0414 & 0.0232 & 0.0235 & 0.0240 & 0.0238 & 0.0310 & 0.0204 & 0.0269 & 0.0267 \\ 
&$\mathrm{\overline{RMSE}}\bm{_\pi}$ & 0.0105 & 0.0065 & 0.0035 & 0.0029 & 0.0084 & 0.0055 & 0.0031 & 0.0026 & 0.0001 & 0.0001 & 0.0001 & 0.0001 & 0.0001 & 0.0001 & 0.0001 & 0.0001 \\ 
&$\mathrm{\overline{MSE}}\bm{_\pi}$  & 0.0001 & 0.0001 & 0.0000 & 0.0000 & 0.0001 & 0.0000 & 0.0000 & 0.0000 & 0.0000 & 0.0000 & 0.0000 & 0.0000 & 0.0000 & 0.0000 & 0.0000 & 0.0000 \\ 
&AAR & 0.8212 & 0.8505 & 0.8627 & 0.8611 & 0.8368 & 0.8561 & 0.8630 & 0.8595 & 0.7179 & 0.7513 & 0.7497 & 0.7492 & 0.7294 & 0.7716 & 0.7652 & 0.7640 \\ 
&PAR2 & 0.8752 & 0.9159 & 0.9290 & 0.9249 & 0.8978 & 0.9257 & 0.9306 & 0.9282 & 0.1747 & 0.2452 & 0.2348 & 0.2349 & 0.2104 & 0.3131 & 0.2860 & 0.2812 \\ 

\hline
\end{tabular}
\begin{tablenotes}
        \footnotesize
         \item \textbf{Note}. In the row “Method”, there are three different levels of $\bm \delta$; the sequential Gibbs samplings with $\bm \delta=\bm{0.01},\bm{0.1}$ and $\bm{1}$.
      \end{tablenotes}
      \end{threeparttable}
\end{center}
\label{HC1}
\end{sidewaystable}%
\begin{sidewaystable}[htp]
\renewcommand\tabcolsep{3.0pt} 
\small
\caption{Parameter recovery for High Dimension with the Correlation $\rho=0.7$}
\begin{center}
\begin{threeparttable}
\begin{tabular}{llllllllllllllllll}
\hline
\multicolumn{2}{r}{Attribute Number}&\multicolumn{8}{c}{$K=7$}&\multicolumn{8}{c}{$K=15$} \\ \cmidrule(lr){3-10}\cmidrule(lr){11-18}
\multicolumn{2}{r}{Sample Size}&\multicolumn{4}{c}{1000}&\multicolumn{4}{c}{2000}&\multicolumn{4}{c}{1000}&\multicolumn{4}{c}{2000}\\ \cmidrule(lr){3-6}\cmidrule(lr){7-10}\cmidrule(lr){11-14}\cmidrule(lr){15-18}
\multicolumn{2}{r}{Method}&$\bm{0.01}$&$\bm {0.1}$&$\bm {1}$&Ind&$\bm{0.01}$&$\bm {0.1}$&$\bm {1}$&Ind&$\bm{0.01}$&$\bm {0.1}$&$\bm{1}$&Ind&$\bm{0.01}$&$\bm{0.1}$&$\bm{1}$&Ind\\
\hline

DINA &$\mathrm{\overline{Bias}}\bm{_g}$ & 0.0079 & 0.0043 & -0.0000 & -0.0117 & 0.0043 & 0.0021 & -0.0002 & -0.0111 & 0.0051 & -0.0176 & -0.0327 & -0.0038 & 0.0094 & -0.0111 & -0.0342 & -0.0019 \\
&$\mathrm{\overline{RMSE}}\bm{_g}$ & 0.0225 & 0.0201 & 0.0188 & 0.0261 & 0.0147 & 0.0136 & 0.0131 & 0.0221 & 0.0234 & 0.0303 & 0.0450 & 0.0334 & 0.0192 & 0.0200 & 0.0407 & 0.0268 \\ 
&$\mathrm{\overline{MSE}}\bm{_g}$ & 0.0005 & 0.0004 & 0.0004 & 0.0008 & 0.0002 & 0.0002 & 0.0002 & 0.0006 & 0.0006 & 0.0010 & 0.0023 & 0.0012 & 0.0004 & 0.0004 & 0.0019 & 0.0008 \\
&$\mathrm{\overline{Bias}}\bm{_s}$ & 0.0053 & 0.0032 & 0.0026 & 0.0090 & 0.0032 & 0.0021 & 0.0016 & 0.0064 & 0.0083 & 0.0136 & 0.0210 & -0.0082 & 0.0040 & 0.0075 & 0.0204 & -0.0125 \\
&$\mathrm{\overline{RMSE}}\bm{_s}$ & 0.0215 & 0.0208 & 0.0207 & 0.0284 & 0.0152 & 0.0149 & 0.0147 & 0.0238 & 0.0271 & 0.0291 & 0.0394 & 0.0471 & 0.0175 & 0.0185 & 0.0307 & 0.0413 \\ 
&$\mathrm{\overline{MSE}}\bm{_s}$ & 0.0005 & 0.0004 & 0.0004 & 0.0008 & 0.0002 & 0.0002 & 0.0002 & 0.0006 & 0.0008 & 0.0009 & 0.0016 & 0.0031 & 0.0003 & 0.0003 & 0.0010 & 0.0030 \\ 
&$\mathrm{\overline{MN}}\bm{_\pi}$ & 0.0313 & 0.0197 & 0.0292 & 0.1556 & 0.0153 & 0.0108 & 0.0168 & 0.1443 & 0.0193 & 0.1174 & 0.1447 & 0.1425 & 0.0368 & 0.0910 & 0.1454 & 0.1449 \\
&$\mathrm{\overline{RMSE}}\bm{_\pi}$ & 0.0040 & 0.0028 & 0.0023 & 0.0036 & 0.0029 & 0.0021 & 0.0016 & 0.0032 & 0.0001 & 0.0001 & 0.0001 & 0.0001 & 0.0001 & 0.0001 & 0.0001 & 0.0001 \\
&$\mathrm{\overline{MSE}}\bm{_\pi}$ & 0.0000 & 0.0000 & 0.0000 & 0.0002 & 0.0000 & 0.0000 & 0.0000 & 0.0002 & 0.0000 & 0.0000 & 0.0000 & 0.0000 & 0.0000 & 0.0000 & 0.0000 & 0.0000 \\
&AAR & 0.9353 & 0.9441 & 0.9488 & 0.8951 & 0.9425 & 0.9471 & 0.9493 & 0.9028 & 0.8834 & 0.8892 & 0.8505 & 0.8091 & 0.8868 & 0.9005 & 0.8642 & 0.8113 \\ 
&PAR2 & 0.9786 & 0.9839 & 0.9862 & 0.9418 & 0.9833 & 0.9856 & 0.9869 & 0.9494 & 0.6998 & 0.7248 & 0.5877 & 0.4968 & 0.7094 & 0.7654 & 0.6348 & 0.5028 \\
GDINA&$\mathrm{\overline{Bias}}\bm{_\lambda}$ & 0.0168 & 0.0262 & 0.0341 & 0.0842 & 0.0051 & 0.0114 & 0.0179 & 0.0914 & 0.0311 & 0.0547 & 0.0697 & 0.0684 & 0.0515 & 0.0482 & 0.0903 & 0.0858 \\ 
&$\mathrm{\overline{RMSE}}\bm{_\lambda}$ & 0.1604 & 0.1477 & 0.1454 & 0.1920 & 0.1440 & 0.1327 & 0.1288 & 0.1942 & 0.1664 & 0.1929 & 0.2139 & 0.2117 & 0.1514 & 0.1906 & 0.2319 & 0.2246 \\ 
&$\mathrm{\overline{MSE}}\bm{_\lambda}$ & 0.0309 & 0.0262 & 0.0251 & 0.0459 & 0.0279 & 0.0233 & 0.0215 & 0.0503 & 0.0323 & 0.0445 & 0.0601 & 0.0591 & 0.0291 & 0.0448 & 0.0760 & 0.0701 \\ 
&$\mathrm{\overline{MN}}\bm{_\pi}$ & 0.0412 & 0.0303 & 0.0515 & 0.1693 & 0.0383 & 0.0238 & 0.0327 & 0.1585 & 0.0559 & 0.1413 & 0.1569 & 0.1562 & 0.0465 & 0.1134 & 0.1422 & 0.1418 \\ 
&$\mathrm{\overline{RMSE}}\bm{_\pi}$ & 0.0074 & 0.0047 & 0.0034 & 0.0058 & 0.0062 & 0.0040 & 0.0027 & 0.0054 & 0.0001 & 0.0001 & 0.0001 & 0.0001 & 0.0001 & 0.0001 & 0.0001 & 0.0001 \\ 
&$\mathrm{\overline{MSE}}\bm{_\pi}$ & 0.0001 & 0.0000 & 0.0000 & 0.0004 & 0.0001 & 0.0000 & 0.0000 & 0.0004 & 0.0000 & 0.0000 & 0.0000 & 0.0000 & 0.0000 & 0.0000 & 0.0000 & 0.0000 \\ 
&AAR & 0.8861 & 0.9046 & 0.9108 & 0.8800 & 0.8866 & 0.9013 & 0.9061 & 0.8740 & 0.8143 & 0.8185 & 0.7807 & 0.7769 & 0.8007 & 0.8242 & 0.7839 & 0.7736 \\ 
&PAR2  & 0.9303 & 0.9559 & 0.9612 & 0.9522 & 0.9323 & 0.9525 & 0.9577 & 0.9432 & 0.4738 & 0.4750 & 0.3406 & 0.3249 & 0.4330 & 0.4921 & 0.3601 & 0.3181 \\ 
\hline
\end{tabular}
\begin{tablenotes}
        \footnotesize
         \item \textbf{Note}. In the row “Method”, there are three different levels of $\bm \delta$; the sequential Gibbs samplings with $\bm \delta=\bm{0.01},\bm{0.1}$ and $\bm{1}$. 

      \end{tablenotes}
      \end{threeparttable}
\end{center}
\label{HC2}
\end{sidewaystable}%

\begin{table}[htp]
\caption{The average computational time}
\begin{center}
\begin{spacing}{1}
\begin{threeparttable}
\begin{tabular}{ccccccccccc}
\hline

&\multicolumn{4}{c}{DINA}&\multicolumn{4}{c}{GDINA}\\\cmidrule(l){2-5}\cmidrule(lr){6-9}
($N,K$)&Sim&Seq&Ind&sSim&Sim&Seq&Ind&sSim\\
\hline
(1000,3) & 388.34 & 9.9296 & 10.809 & 9.6042 &224.01 & 52.096 & 53.002 & 27.426\\
(1000,5) &591.50 & 10.671 & 12.066 & 24.890 & 370.68 & 50.496 & 52.038 & 49.066\\
(1000,7)& --- & 11.178 & 13.082 & 84.437& --- & 51.232 & 53.165& 133.16\\
(1000,15) & --- & 19.895 & 17.347 & ---& --- & 64.106 & 54.538&---\\
(2000,3) & 856.37 & 19.573 & 21.568 & 18.986 &516.98 & 103.63 & 105.93 & 53.263\\
(2000,5) &1308.2 & 20.877 & 23.815 & 48.351 & 828.19 & 99.263& 102.31 & 94.644\\
(2000,7)& --- & 21.753 & 25.917 & 165.23 & ---  & 99.180 & 104.39 & 252.84\\
(2000,15)& --- & 32.277 & 33.705& ---  & --- & 106.66 & 101.44 &---\\
\hline
\end{tabular}
\begin{tablenotes}
        \footnotesize
        \item \textbf{Note}. The ``Sim", ``Seq", ``Ind" and ``sSim" represent the simultaneous (implemented by JAGS), sequential, independent and self-compiled simultaneous Gibbs sampling methods, respectively. The time is reported in seconds.
        \end{tablenotes}
      \end{threeparttable}
\end{spacing}
\end{center}
\label{Time}
\end{table}%

 

\section{Real Data Analysis} \label{6}


In this analysis, the DINA and GDINA models are used to deal with the Tatsuoka's fraction-subtraction data \citep{Tatsuoka2002analysis}. 
The fraction-subtraction data has been widely analyzed. For the data set, the $Q$-matrix \citep{de2004higher} and contents are shown in Table \ref{Frac_Sub}. This data set contains responses of $536$ middle school students (i.e., $N=536$) to $20$ items (i.e., $J=20$). There are 8 attributes and $2^8=256$ latent classes.

\begin{table}[htp]
\caption{The $Q$-matrix and Items for Fractions-Subtraction Data}
\begin{center}
\begin{spacing}{.9}
\begin{tabular}{llcccccccccc}
\hline
ID&Item&$\alpha_1$&$\alpha_2$&$\alpha_3$&$\alpha_4$&$\alpha_5$&$\alpha_6$&$\alpha_7$&$\alpha_8$\\[4pt]
\hline
1&$\frac{5}{3}-\frac{3}{4}$ & 0 & 0 & 0 & 1 & 0 & 1 & 1 & 0 \\[4pt]
2&$\frac{3}{4}-\frac{3}{8}$ & 0 & 0 & 0 & 1 & 0 & 0 & 1 & 0 \\ [4pt]
3&$\frac{5}{6}-\frac{1}{9}$& 0 & 0 & 0 & 1 & 0 & 0 & 1 & 0 \\ [4pt]
4&$3\frac{1}{2}-2\frac{3}{2}$ & 0 & 1 & 1 & 0 & 1 & 0 & 1 & 0 \\ [4pt]
5&$4\frac{3}{5}-3\frac{4}{10}$& 0 & 1 & 0 & 1 & 0 & 0 & 1 & 1 \\ [4pt]
6&$\frac{6}{7}-\frac{4}{7}$ & 0 & 0 & 0 & 0 & 0 & 0 & 1 & 0 \\ [4pt]
7&$3-2\frac{1}{5}$ & 1 & 1 & 0 & 0 & 0 & 0 & 1 & 0 \\ [4pt]
8& $\frac{2}{3}-\frac{2}{3}$& 0 & 0 & 0 & 0 & 0 & 0 & 1 & 0 \\ [4pt]
9& $3\frac{7}{8}-2$& 0 & 1 & 0 & 0 & 0 & 0 & 0 & 0 \\ [4pt]
10&$4\frac{4}{12}-2\frac{7}{12}$ & 0 & 1 & 0 & 0 & 1 & 0 & 1 & 1 \\ [4pt]
11&$4\frac{1}{3}-2\frac{4}{3}$ & 0 & 1 & 0 & 0 & 1 & 0 & 1 & 0 \\ [4pt]
12&$\frac{11}{8}-\frac{1}{8}$ & 0 & 0 & 0 & 0 & 0 & 0 & 1 & 1 \\ [4pt]
13&$3\frac{3}{8}-2\frac{6}{5}$ & 0 & 1 & 0 & 1 & 1 & 0 & 1 & 0 \\ [4pt]
14&$3\frac{4}{5}-3\frac{2}{5}$ & 0 & 1 & 0 & 0 & 0 & 0 & 1 & 0 \\ [4pt]
15&$2-\frac{1}{3}$ & 1 & 0 & 0 & 0 & 0 & 0 & 1 & 0 \\ [4pt]
16&$4\frac{5}{7}-1\frac{4}{7}$ & 0 & 1 & 0 & 0 & 0 & 0 & 1 & 0 \\ [4pt]
17&$7\frac{3}{5}-2\frac{4}{5}$ & 0 & 1 & 0 & 0 & 1 & 0 & 1 & 0 \\ [4pt]
18&$4\frac{1}{10}-2\frac{8}{10}$ & 0 & 1 & 0 & 0 & 1 & 1 & 1 & 0 \\ [4pt]
19&$4-1\frac{4}{3}$ & 1 & 1 & 1 & 0 & 1 & 0 & 1 & 0 \\ [4pt]
20&$4\frac{1}{3}-1\frac{5}{3}$ & 0 & 1 & 1 & 0 & 1 & 0 & 1 & 0 \\ [4pt]
  \hline
\end{tabular}
\end{spacing}
\end{center}
\label{Frac_Sub}
\end{table}%

\begin{table}[htp]
\caption{Summary of analyzing conditions for the Fraction-Subtraction Data}
\begin{center}
\begin{spacing}{1}
\begin{tabular}{ccc}
\hline
&DINA&GDINA\\
\hline
$\mbox{GDINA } \bm\lambda$ prior & ---&N(0,1)\\
$\mbox{DINA } \bm s,\bm g$ prior &  Beta(1,1)&---\\
$\bm\delta$&$\bm{0.1}$ &$\bm{0.1}$  \\
Chain Length&2000&4000\\
Burn-in&1000&2000\\
\hline
\end{tabular}
\end{spacing}
\end{center}
\label{datasetting1}
\end{table}%

We fit both the DINA and GDINA models. Based on the results of simulation studies, we set the hyper-parameter of the Dirichlet priror $\bm \delta=\bm{0.1}$. When applying the GDINA model, we assume the item parameter prior for all items as $\lambda^{(w)} \sim N(0,1) $ for both $w=0$ and $w>0$.
The prior hyper-parameters and MCMC chain lengths and burn-in are listed in Table \ref{datasetting1}. The computation of analyses is performed by a 2018 MacBook Pro with 2.2 GHz Intel Core i7, 16 GB 2400 MHz DDR4 and Radeon Pro 555X 4096 MB; Intel UHD Graphics 630 1536 MB. The only feasible case for JAGS is running the 8-attribute DINA model with the simultaneous Gibbs sampling. 
For the DINA model three methods, the independent, sequential and simultaneous Gibbs sampling by implemented JAGS, are compared. However, for the GDINA model, the independent and sequential Gibbs sampling algorithms are used to compare.

\begin{figure}[htp]
\centering
\subfigure[DINA guessing parameters]{
\label{F_DINA_g}
\includegraphics[width=0.31\linewidth]{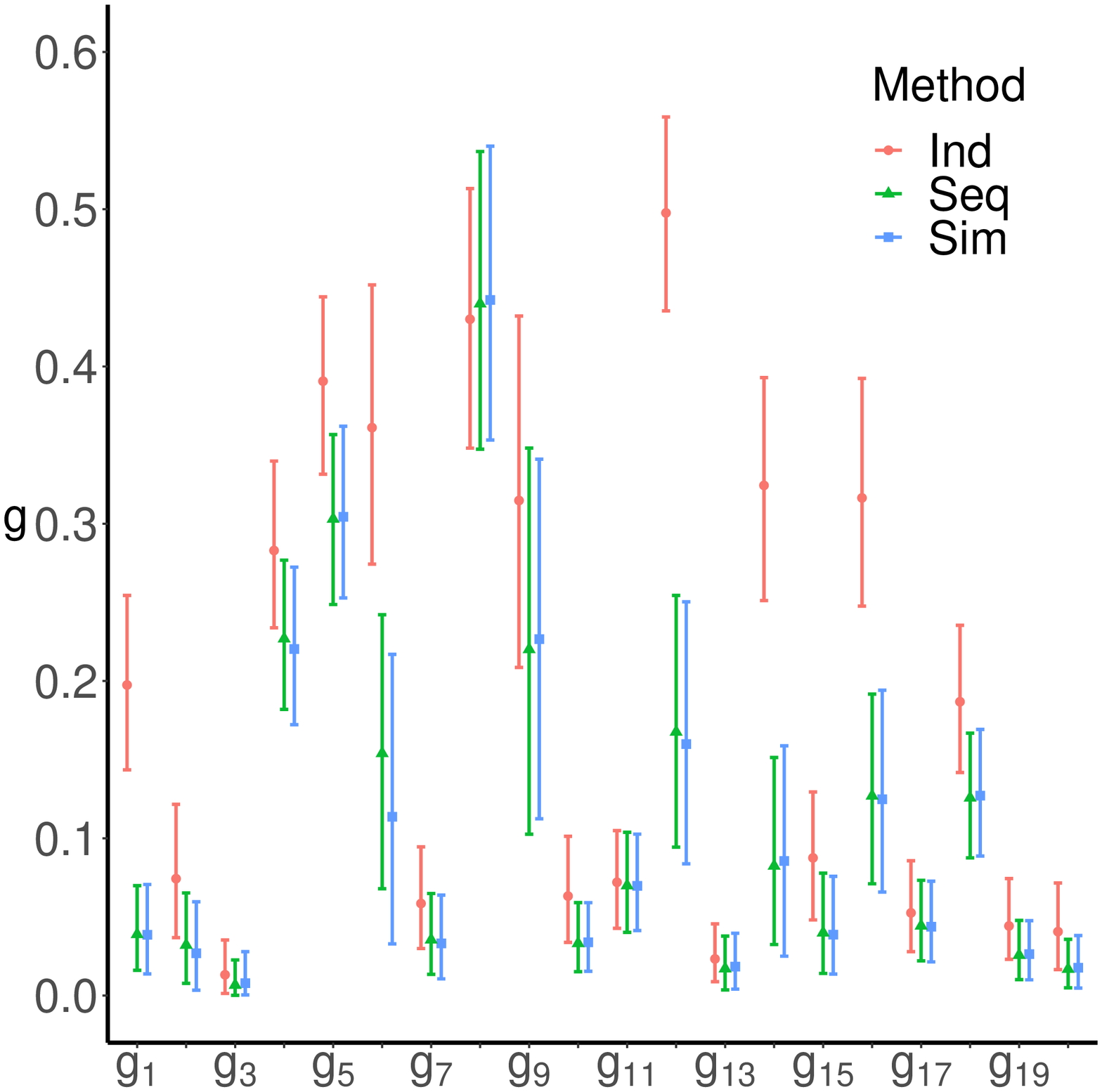}
}%
\vspace{.005\linewidth}
\subfigure[DINA slipping parameters]{
\label{F_DINA_s}
\includegraphics[width=.31\linewidth]{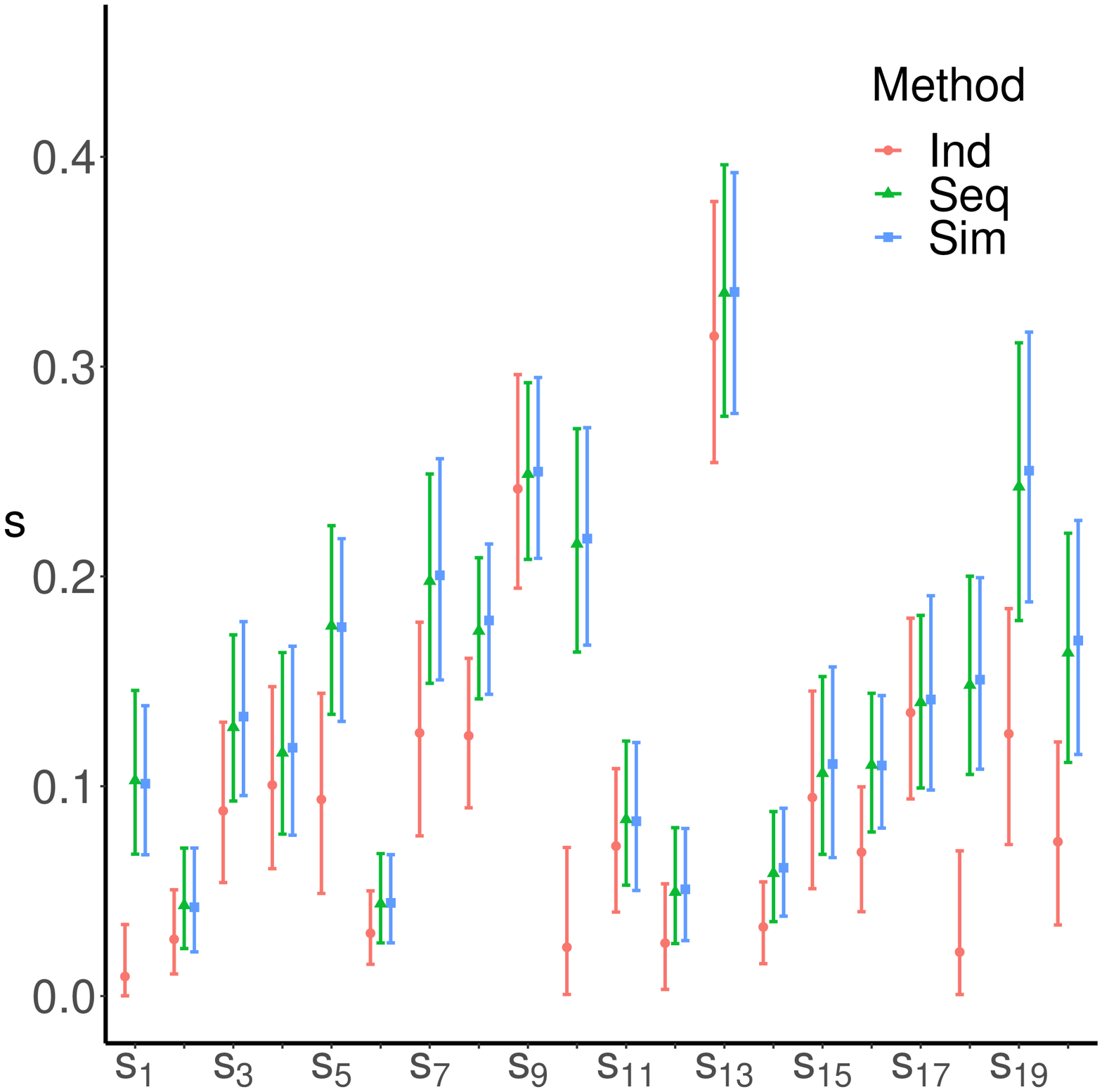}
}%
\vspace{.005\linewidth}
\subfigure[GDINA parameters] {
\label{F_GDINA}
\includegraphics[width=.31\linewidth]{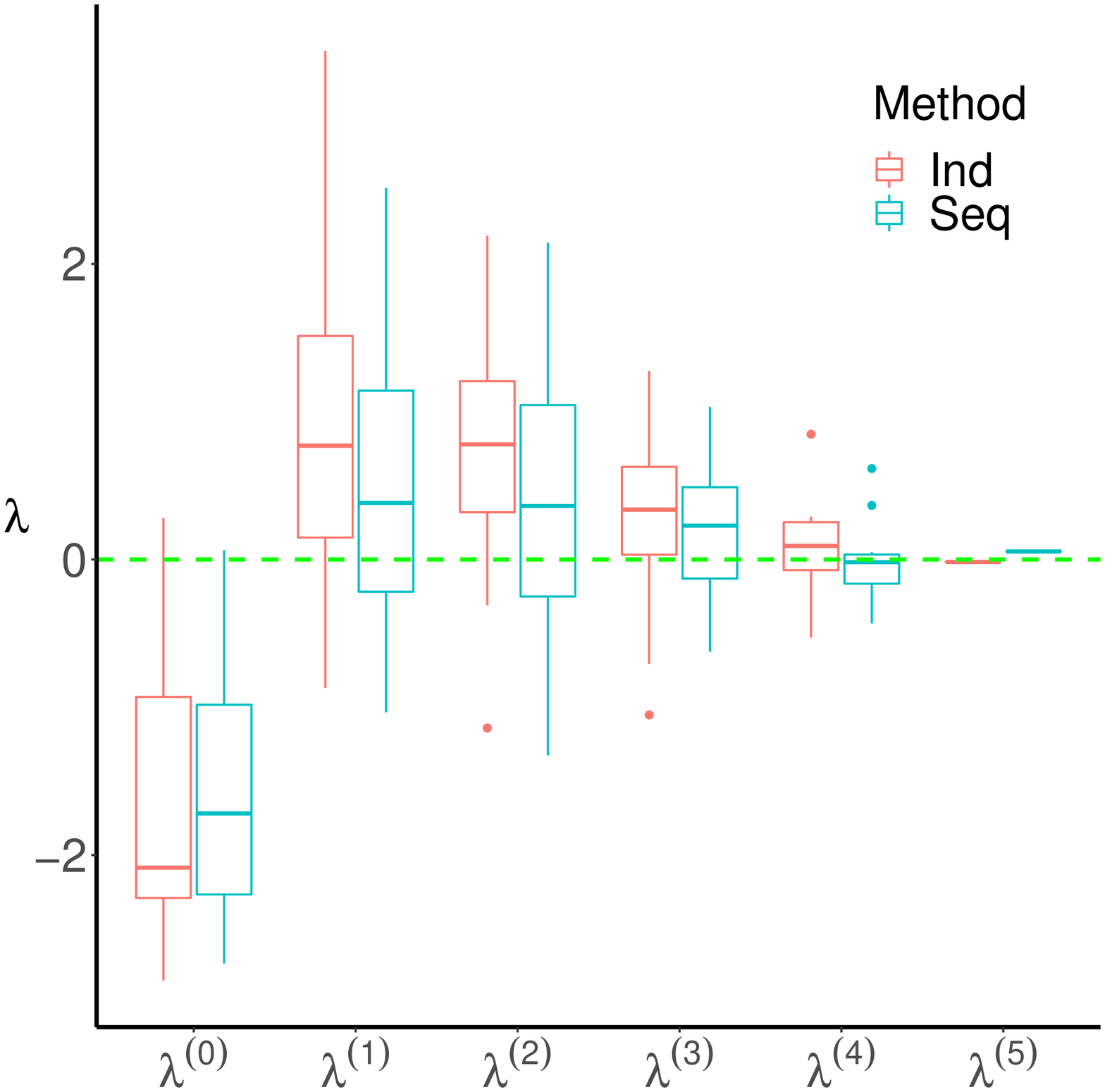}
}%
\centering
\caption{ The estimations of item parameters for the fraction-subtraction data. {The ``Ind", ``Seq” and ``Sim" represent the results from the independent, sequential and simultaneous (by implemented JAGS) Gibbs sampling}, respectively.}
\label{Frac_Item_DINA}
\end{figure}

When to analyze the fraction-subtraction data with the DINA model, the independent, sequential and simultaneous Gibbs sampling methods spend 5.46, 4.91 and 942  seconds, respectively. The off-the-shelf software JAGS is treated as a benchmark. We find that the simultaneous Gibbs sampling using JAGS is time-consuming. 
Figures \ref{F_DINA_g} and \ref{F_DINA_s} show the posterior means and 95\% confidence interval of the guessing and slipping parameters, respectively. When to compare estimation accuracy of item parameters, no matter from posterior means or 95\% confidence interval, the simultaneous Gibbs sampling gives very similar results to the sequential Gibbs sampling, while the latter has obvious speed advantage. Compared with the MCMC estimates obtained by JAGS, the independent Gibbs sampling trends to overestimate the guessing parameters and underestimate the slipping parameters, which may be due to the correlation of the attributes. 


In the GDINA model, the item parameters are intercept and interaction parameters rather than guessing and slipping parameters. The time costs of  the independent and sequential Gibbs sampling algorithms are 19.25 and 17.65 seconds, respectively. The notation $\lambda^{(w)}$ represents the $w$-way interaction. Since the items in the fraction-subtraction data need up to 5 attributes, there exist up to the 5-way interaction parameters. 
The Figure \ref{F_GDINA} presents the box-plot of $w$ versus $\lambda^{(w)}$ for the estimated item parameters. The common property shared by the same-way interactions can be obtained by the box-plot. The item parameter estimations show that the means of intercept parameters and 4-way interaction parameters are negative and the others are positive. 
The conclusion that the intercept term is negative is consistent with our intuition, because   a subject without any required attributes is usually  expected to  have a low positive response probability. 

\section{Discussion} \label{7}
In practice, one computational challenge is that when the number of attributes is large, the existing MCMC for the CDMs may become slow. In this paper, a computationally efficient algorithm, named as the sequential Gibbs sampling, was proposed for a general CDM, i.e., the GDINA model. 
In the situation with small $K$, compared to the existing method (e.g., JAGS), the proposed method can also yield similar results. The proposed method still works well and fast for the case with large $K$.
When $K=15,\ I=2000,\ J=40$ and the model is the GDINA  model, running 3,000 iterations only needs less 110 seconds. Especially, for a large $K$, the computational advantage becomes more significant compared with the simultaneous Gibbs sampling method.  
The proposed method can be easily applied to other CDMs.
In the appendix, we show the algorithm for the DINA model for an illustration.  

In this paper, we only focus on the computational challenge for large $K$, given the $Q$-matrix is correctly specified. 
Most references about identification theory pointed out that the $Q$-matrix need to contain an identity matrix at least
for strict identifiability \citep{xu2018identifying,gu2018partial}.
In practice, however, the $Q$-matrix may be misspecified, and it would be needed to estimate the $Q$-matrix together with the model parameters and latent attributes. The estimation of the $Q$-matrix is known to be a challenging issue, especially when $K$ is large, and  the proposed algorithm may be extended to such applications to help reduce the computational cost of the convensional MCMC approaches.
Another interesting extension is to use the idea in this paper to solve other latent variable modeling problems with many latent attributes. 
Not only for the discrete but also for continuous abilities, this idea may be helpful. 


\bibliographystyle{apacite}
\bibliography{mybib}
\begin{appendix}
\vspace{1cm}

\newpage

\centerline{\Large \textbf{Appendix}}
\noindent\rule{\textwidth}{0.5mm}

\setcounter{equation}{0}
\renewcommand{\theequation}{\Alph{section}.\arabic{equation}}
\renewcommand{\thesection}{\Alph{section}}
\section{Sequential Gibbs sampling for the DINA model}\label{SEQDINA}
Under the DINA model, the item parameters are guessing parameter $\bm g$ and slipping parameter $\bm s$. 
The priors of guessing and slipping parameters follow $\mathrm{Beta}(a_g,b_g)$ and $\mathrm{Beta}(a_s,b_s)$, respectively. The rest of the measurement model remains similarly to the GDINA model. 
The sampling methods for $\bm g, \bm s $ and $\bm \pi$ can be founded in \cite{culpepper2015bayesian}. We only focus on sampling attribute profiles: for fixed $\bm Y,\bm \alpha_{i \backslash k},\bm g, \bm s$ and $\bm \pi$, the full conditional distribution for $\alpha_{ik}$ is proportional to 
\begin{small}
\begin{align}
&p(\alpha_{ik}|\bm Y_i,\bm \alpha_{i \backslash k},\bm g,\bm s,\bm \pi) \notag\\
 \propto&\prod_{j=1}^J \left [P_j(\bm \alpha_i)^{Y_{ij}}(1-P_j(\bm \alpha_i) )^{1-Y_{ij}} \right ]p(\alpha_{ik}|p_{ik})\label{d}\\
\propto&\prod_{j=1}^J \left [(g_j^{(1-\eta_{ij})} (1-s_j)^{\eta_{ij}})^{Y_{ij}} ((1-g_j)^{(1-\eta_{ij})} s_j^{\eta_{ij}})^{1-Y_{ij}} \right ]p_{ik}^{\alpha_{ik}} (1-p_{ik})^{1-
\alpha_{ik}}\\
\propto& \prod_{j=1}^J \left [ (s_j^{1-Y_{ij}} (1-s_j)^{Y_{ij}}) ^ {\eta_{ij}} (g_j^{Y_{ij}} (1-g_j)^{1-Y_{ij}}) ^{1-\eta_{ij}}\right ]p_{ik}^{\alpha_{ik}} (1-p_{ik})^{1-
\alpha_{ik}}\label{o}\\
\propto& \prod_{j=1}^J \left [ (s_j^{1-Y_{ij}} (1-s_j)^{Y_{ij}}) ^ {\prod_{k=1}^K \alpha_{ik}^{q_{jk}}} (g_j^{Y_{ij}} (1-g_j)^{1-Y_{ij}}) ^{1-\prod_{k=1}^K \alpha_{ik}
^{q_{jk}}}\right ]p_{ik}^{\alpha_{ik}} (1-p_{ik})^{1-\alpha_{ik}},
\end{align}
\end{small}
\noindent where $p_{ik}$ is the prior conditional probability as in   Equation (\ref{seq2prior}). For simplify, we define a set $\Omega_{ik}=\{j \mid  \prod_{k' \neq k}\alpha_{ik'}^{q_{jk'}}=1 \& q_{jk}=1\}$, and the complementary set is $\Omega_{ik}^c=\{ {j \mid  j\notin 
\Omega_{ik} \& j= 1,\dots,J} \}$. We   know that when item $j$ belongs to $\Omega_{ik}$, $\eta_{ij} =\alpha_{ik}$. The items in the set $\Omega_{ik}^c$ satisfy at least one of two conditions, $\prod_{k' \neq k}\alpha_{ik'}^{q_{jk'}}=0$ and ${q_{jk}}=0$, which indicate that the value of $\alpha_{ik}$ doesn't affect the value of $\eta_{ij}$. In other words, the items in the set $\Omega_{ik}^c$ don't affect the full conditional distribution. We can rewrite the full conditional distribution as follows:
\begin{equation}
\begin{aligned}\label{seqDINAalpha}
& p(\alpha_{ik}|\bm Y_i,\bm \alpha_{i \backslash k},\bm g,\bm s ,\bm p) \\
\propto & \prod_{j \in \Omega_{ik} }\left [ (s_j^{1-Y_{ij}} (1-s_j)^{Y_{ij}}) ^ {\alpha_{ik}} (g_j^{Y_{ij}} (1-g_j)^{1-Y_{ij}}) ^{1-\alpha_{ik}}\right ] \\
& \prod_{j \notin \Omega_{ik} }\left [ (s_j^{1-Y_{ij}} (1-s_j)^{Y_{ij}}) ^ 0 (g_j^{Y_{ij}} (1-g_j)^{1-Y_{ij}}) ^{1}\right ]p_{ik}^{\alpha_{ik}} (1-p_{ik})^{1-\alpha_{ik}}\\
\propto & \prod_{j \in \Omega_{ik} }\left [ (s_j^{1-Y_{ij}} (1-s_j)^{Y_{ij}}) ^ {\alpha_{ik}} (g_j^{Y_{ij}} (1-g_j)^{1-Y_{ij}}) ^{1-\alpha_{ik}}\right ] p_{ik}^{\alpha_{ik}} 
(1-p_{ik})^{1-\alpha_{ik}}.
\end{aligned}
\end{equation}
Obviously, the posterior distribution of $\alpha_{ik}$ is a Bernoulli distribution $\mathrm{Bernoulli} (\hat{p}_{ik})$ with the parameter
\begin{small}
\begin{equation}
\nonumber 
 \hat{p}_{ik} =\frac {\prod_{j \in \Omega_{ik} }\left [ s_j^{1-Y_{ij}} (1-s_j)^{Y_{ij}}\right ] p_{ik}} {\prod_{j \in \Omega_{ik} }\left [ s_j^{1-Y_{ij}} (1-s_j)^{Y_{ij}}\right ] p_{ik} +\prod_{j \in \Omega_{ik}} \left [(g_j^{Y_{ij}} (1-g_j)^{1-Y_{ij}})\right] (1-p_{ik})}.
\end{equation}
\end{small}

We next introduce the initial values for the DINA model. The initial values of item parameters $ \bm g^{(0)}, \bm s^{(0)}$ are randomly sampled from the uniform distribution $U[0, 0.4]$. The initial population parameter $\bm\pi$ is the $C$-dimensional $\bm{1/C}$; the hyper-parameters are $a_s=b_s=a_g=b_g=1$. The other initial values are same as those in section \ref{6.1}. The sequential Gibbs sampling for the DINA model is presented in Algorithm \ref{A2}.

\begin{algorithm}[htp]
 \label{A2}
            \caption{Sequential Gibbs Sampling for DINA}
            \KwIn{Initialize $\bm g^{(0)},\bm s^{(0)},\bm\alpha^{(0)},\bm\pi^{(0)},\bm Y,m=0, M$ and specify priors.} 
            \KwOut{Markov chains of $\bm g^{}, \bm s^{},\bm\alpha,\bm\pi$. }

            \While{$m < M$}{
            Sample attribute profiles from Equation (\ref{seqDINAalpha}). \\
            Sample the other parameters according to the reference \citep{culpepper2015bayesian}.\\
            Set $m= m+1$.
            }
\end{algorithm}
%

\setcounter{equation}{0}
\section{Full conditional distribution for $\lambda_j$} \label{colsed form}
Focusing on the $j$th item, we can get a classical linear regression as follows $$\bm Z_{j} = \bm X_j \bm \lambda_j + \bm\varepsilon_j,$$
where $\bm \lambda_j $ is the item parameter and the residual $\bm\varepsilon_j=(\varepsilon_1,\varepsilon_2,\cdots,\varepsilon_N)'$ is a random sample from standard normal distribution. The kernel of likelihood function is given as
\begin{equation}
p(\bm Z_j|\bm X_j,\bm\lambda_j) \propto \exp \left(-\frac{1}{2}(\bm Z_j-\bm X_j \bm \lambda_j) 'I^{-1} (\bm Z_j-\bm X_j \bm \lambda_j) \right),
\end{equation}
where the $I$ represents an identity matrix. Assuming the joint prior for $\bm\lambda_j$ is $N(\bm \mu_{\lambda_j},\bm\Sigma_{\lambda_j})$, the specific form of parameter's prior is 
\begin{equation}
p(\bm\lambda_j) \propto \exp \left(-\frac{1}{2}(\bm \lambda_j-\bm \mu_{\lambda_j}) '\bm\Sigma_{\lambda_j}^{-1}(\bm \lambda_j-\bm \mu_{\lambda_j}) \right)
\end{equation}
According to   the Bayesian linear regression, the kernel of the posterior is 
\begin{equation}
\begin{aligned}
&p(\bm\lambda_j|\bm Z_j,\bm X_j, \bm \lambda_j) \\
\propto& \exp \left [-\frac{1}{2} \left((\bm Z_j-\bm X_j \bm \lambda_j) 'I^{-1} (\bm Z_j-\bm X_j \bm \lambda_j) +(\bm \lambda_j-\bm \mu_{\lambda_j}) '\bm\Sigma_{\lambda_j}^{-1}(\bm \lambda_j-\bm \mu_{\lambda_j})\right) \right]\\
\propto  & \exp \left [-\frac{1}{2} \left( \bm\lambda_j'(\bm X_j'\bm X_j+\bm\Sigma_{\lambda_j}^{-1})\bm\lambda_j- \bm \lambda_j'(\bm X_j'\bm I^{-1} \bm Z_j+\bm\Sigma_{\lambda_j}^{-1} \bm \mu_{\lambda_j})-(\bm X_j'\bm I^{-1} \bm Z_j+\bm\Sigma_{\lambda_j}^{-1} \bm \mu_{\lambda_j})' \bm\lambda_j   \right) \right].
\end{aligned}
\label{Kernel}
\end{equation}
 Let $\hat {\bm\Sigma}_{\lambda_j}^{-1}=\bm X_j \bm X_j'+\bm \Sigma_{\lambda_j}^{-1}$, using the undetermined coefficient method to solve $\hat {\bm\mu}_{\lambda_j}=\hat{\bm \Sigma}_{\lambda_j} (\bm X_j '\bm Z_{ j}+\bm\Sigma_{\lambda_j}^{-1}\bm\mu_{\lambda_j} )$, so we can get the full conditional distribution for $j$th item's parameters easily. The full conditional distribution is $N(\hat {\bm\mu}_{\lambda_j},\hat{\bm \Sigma}_{\lambda_j})$.
  

\section{Sample from the truncated multivariate normal distribution}\label{sampling}
Because the posterior is a normal distribution, only the multivariate normal distribution with truncation needs to de derived. Assuming that a random vector $ X \sim N_p(\bm \mu,\bm\Sigma)$ and $ X =(X_1,X_2,\cdots,X_p)'$ is a $p$-dimensional vector, where$$\bm\mu=(\mu_1,\cdots, \mu_p)',$$$$\bm\Sigma=\begin{pmatrix}
\sigma_{11}&\sigma_{12}&\cdots&\sigma_{1p}\\
\sigma_{21}&\sigma_{22}&\cdots&\sigma_{2p}\\
\vdots&\vdots&\ddots&\vdots\\
\sigma_{p1}&\sigma_{p2}&\cdots&\sigma_{pp}
\end{pmatrix}.$$

When we ignore the truncation, the marginal distribution of $X_1$ is a normal distribution which can be used to generate $x_1$, and then the conditional distributions of the following parameters: $$X_2|X_1=x_1,$$ $$\vdots$$ $$X_p|X_1=x_1, \cdots,X_{p-1}=x_{p-1},$$ are also normal distributions \citep{anderson1962introduction} which are used to generate $x_2,\cdots,x_p$. As a result, we can get a realization $(x_1,x_2,\cdots,x_p)'$ following $ N_p(\bm \mu,\bm\Sigma)$.

In this paper, the first component $X_1$ is negative, the else components are positive. Imposing the restrictions to $(x_1,x_2,\cdots,x_p)'$, we sample $x_1$ from the interval $(-\infty, 0)$ part of the marginal distribution of $X_1$ and sample $x_2,\cdots,x_{p}$ from the $(0, \infty)$ of remaining conditional distributions. 

Furthermore, through this method, more complex restrictions can be easy to impose. For any $X_i$, the left censoring, right censoring and interval censoring can be employed. Generating censoring data from a unidimensional normal distribution is easy, so this method is rather flexible and simple.

\section{The $Q$-matrices for different $K$ }\label{Qmatrix}
The $Q$-matrices are used for different $K$ in the simulation studies, please see Table \ref{Q-matrix}.
\begin{table}[htp]
\scriptsize
\setlength{\tabcolsep}{0.45mm}
\begin{tabular}{llccccccccccccccccccccccccccccccccccccccccc}
\hline
&A&1&2&3&4&5&6&7&8&9&10&11&12&13&14&15&16&17&18&19&20&21&22&23&24&25&26&27&28&29&30&31&32&33&34&35&36&37&38&39&40\\
\hline
\multirow{3}{*}{K=3}&A1&1&0&0&1&0&0&1&1&0&1&1&1&1&1&1&1&1&0&0&1&0&0&1&1&0&1&1&1&1&1&1&1&1&0&0&1&0&0&1&1\\
&A2&0&1&0&0&1&0&1&0&1&1&0&1&1&1&1&1&0&1&0&0&1&0&1&0&1&1&0&1&1&1&1&1&0&1&0&0&1&0&1&0\\
&A3&0&0&1&0&0&1&0&1&1&0&1&1&1&1&1&1&0&0&1&0&0&1&0&1&1&0&1&1&1&1&1&1&0&0&1&0&0&1&0&1\\
\hline
\multirow{5}{*}{K=5}&A1&1&0&0&0&0&1&0&0&0&0&1&0&0&0&1&1&0&0&1&1&1&0&0&0&0&1&0&0&0&1&1&0&0&1&1&1&0&0&0&0\\
&A2&0&1&0&0&0&0&1&0&0&0&1&1&0&0&0&1&1&0&0&1&0&1&0&0&0&1&1&0&0&0&1&1&0&0&1&0&1&0&0&0\\
&A3&0&0&1&0&0&0&0&1&0&0&0&1&1&0&0&1&1&1&0&0&0&0&1&0&0&0&1&1&0&0&1&1&1&0&0&0&0&1&0&0\\
&A4&0&0&0&1&0&0&0&0&1&0&0&0&1&1&0&0&1&1&1&0&0&0&0&1&0&0&0&1&1&0&0&1&1&1&0&0&0&0&1&0\\
&A5&0&0&0&0&1&0&0&0&0&1&0&0&0&1&1&0&0&1&1&1&0&0&0&0&1&0&0&0&1&1&0&0&1&1&1&0&0&0&0&1\\
\hline
\multirow{7}{*}{K=7}&A1&1&0&0&0&0&0&0&1&0&0&0&0&0&0&1&0&0&0&0&1&1&0&0&0&1&0&0&0&0&0&0&1&0&0&0&0&1&1&0&0\\
&A2&0&1&0&0&0&0&0&0&1&0&0&0&0&0&1&1&0&0&0&0&1&0&0&0&0&1&0&0&0&0&0&1&1&0&0&0&0&1&0&0\\
&A3&0&0&1&0&0&0&0&0&0&1&0&0&0&0&0&1&1&0&0&0&0&1&0&0&0&0&1&0&0&0&0&0&1&1&0&0&0&0&1&0\\
&A4&0&0&0&1&0&0&0&0&0&0&1&0&0&0&0&0&1&1&0&0&0&1&1&0&0&0&0&1&0&0&0&0&0&1&1&0&0&0&1&1\\
&A5&0&0&0&0&1&0&0&0&0&0&0&1&0&0&0&0&0&1&1&0&0&1&1&1&0&0&0&0&1&0&0&0&0&0&1&1&0&0&1&1\\
&A6&0&0&0&0&0&1&0&0&0&0&0&0&1&0&0&0&0&0&1&1&0&0&1&1&0&0&0&0&0&1&0&0&0&0&0&1&1&0&0&1\\
&A7&0&0&0&0&0&0&1&0&0&0&0&0&0&1&0&0&0&0&0&1&1&0&0&1&0&0&0&0&0&0&1&0&0&0&0&0&1&1&0&0\\
\hline
\multirow{15}{*}{K=15}&A1&1&0&0&0&0&0&0&0&0&0&0&0&0&0&0&1&0&0&0&0&0&0&0&0&0&0&0&0&0&0&1&0&0&0&0&1&0&0&0&0\\
&A2&0&1&0&0&0&0&0&0&0&0&0&0&0&0&0&0&1&0&0&0&0&0&0&0&0&0&0&0&0&0&1&0&0&0&0&1&0&0&0&0\\
&A3&0&0&1&0&0&0&0&0&0&0&0&0&0&0&0&0&0&1&0&0&0&0&0&0&0&0&0&0&0&0&0&1&0&0&0&1&0&0&0&0\\
&A4&0&0&0&1&0&0&0&0&0&0&0&0&0&0&0&0&0&0&1&0&0&0&0&0&0&0&0&0&0&0&0&1&0&0&0&0&1&0&0&0\\
&A5&0&0&0&0&1&0&0&0&0&0&0&0&0&0&0&0&0&0&0&1&0&0&0&0&0&0&0&0&0&0&0&0&1&0&0&0&1&0&0&0\\
&A6&0&0&0&0&0&1&0&0&0&0&0&0&0&0&0&0&0&0&0&0&1&0&0&0&0&0&0&0&0&0&0&0&1&0&0&0&1&0&0&0\\
&A7&0&0&0&0&0&0&1&0&0&0&0&0&0&0&0&0&0&0&0&0&0&1&0&0&0&0&0&0&0&0&0&0&0&1&0&0&0&1&0&0\\
&A8&0&0&0&0&0&0&0&1&0&0&0&0&0&0&0&0&0&0&0&0&0&0&1&0&0&0&0&0&0&0&0&0&0&1&0&0&0&1&0&0\\
&A9&0&0&0&0&0&0&0&0&1&0&0&0&0&0&0&0&0&0&0&0&0&0&0&1&0&0&0&0&0&0&0&0&0&0&1&0&0&1&0&0\\
&A10&0&0&0&0&0&0&0&0&0&1&0&0&0&0&0&0&0&0&0&0&0&0&0&0&1&0&0&0&0&0&0&0&0&0&1&0&0&0&1&0\\
&A11&0&0&0&0&0&0&0&0&0&0&1&0&0&0&0&0&0&0&0&0&0&0&0&0&0&1&0&0&0&0&0&0&0&0&0&0&0&0&1&0\\
&A12&0&0&0&0&0&0&0&0&0&0&0&1&0&0&0&0&0&0&0&0&0&0&0&0&0&0&1&0&0&0&0&0&0&0&0&0&0&0&1&0\\
&A13&0&0&0&0&0&0&0&0&0&0&0&0&1&0&0&0&0&0&0&0&0&0&0&0&0&0&0&1&0&0&0&0&0&0&0&0&0&0&0&1\\
&A14&0&0&0&0&0&0&0&0&0&0&0&0&0&1&0&0&0&0&0&0&0&0&0&0&0&0&0&0&1&0&0&0&0&0&0&0&0&0&0&1\\
&A15&0&0&0&0&0&0&0&0&0&0&0&0&0&0&1&0&0&0&0&0&0&0&0&0&0&0&0&0&0&1&0&0&0&0&0&0&0&0&0&1\\
\hline
\end{tabular}
\caption{The transposed $Q$-matrices for different $K$}
\label{Q-matrix}
\end{table}

\section{Analyses of TIMSS 2007}

The Trends in Mathematics and Science Study (TIMSS), a quadrennial assessment, assessed the mathematics and science abilities of fourth and eighth students since 1995. TIMSS 2007 (Grade 4) dataset with 25 mathematics (dichotomized) items used in \cite{lee2011cognitive}, \cite{park2014extension} and \cite{park2018explanatory}. The dataset includes a sample of 698 Austrian students. The chosen data contain $J=$ 25 items (i.e., booklets 4 and 5), which consist of two parts: 11 items released for booklet 4 were new items developed for TIMSS 2007 and the remaining 14 items from booklet 5 were previously administered during TIMSS 2003. 
\begin{table}[htp]
\centering
\begin{spacing}{.9}
\begin{tabular}{cccccccccccccccc}
\toprule
 & \multicolumn{8}{c}{Number} &\multicolumn{4}{c}{\tabincell{c}{Geometric Shape\\and Measure}} & \multicolumn{3}{c}{Data Display} \\  \cmidrule(lr){2-9} \cmidrule(lr){10-13} \cmidrule(lr){14-16}
Item-ID &1&2&3&4&5&6&7&8&9&10&11&12&13&14&15\\
\bottomrule
  M041052 & 1 & 1 & 0 & 0 & 0 & 0 & 0 & 0 & 0 & 0 & 0 & 0 & 0 & 0 &  0 \\ 
  M041056 & 0 & 0 & 0 & 0 & 1 & 0 & 0 & 0 & 0 & 0 & 0 & 0 & 0 & 0 &  0 \\ 
  M041069 & 0 & 1 & 0 & 1 & 1 & 0 & 0 & 0 & 0 & 0 & 0 & 0 & 0 & 0 &  0 \\ 
  M041076 & 0 & 0 & 1 & 0 & 0 & 1 & 0 & 0 & 0 & 0 & 0 & 0 & 0 & 0 &  0 \\ 
  M041281 & 0 & 1 & 1 & 0 & 0 & 0 & 0 & 1 & 0 & 0 & 0 & 0 & 0 & 0 &  0 \\ 
  M041164 & 0 & 0 & 0 & 0 & 0 & 0 & 0 & 0 & 0 & 1 & 0 & 1 & 0 & 0 &  0 \\ 
  M041146 & 0 & 0 & 0 & 0 & 0 & 0 & 0 & 0 & 1 & 1 & 0 & 1 & 0 & 0 &  0 \\ 
  M041152 & 1 & 1 & 1 & 0 & 0 & 0 & 0 & 0 & 0 & 1 & 1 & 0 & 0 & 0 &  0 \\ 
  M041258A & 0 & 0 & 0 & 0 & 0 & 0 & 0 & 0 & 0 & 1 & 0 & 0 & 0 & 0 &  0 \\ 
  M041258B & 0 & 0 & 0 & 0 & 0 & 0 & 0 & 0 & 1 & 1 & 0 & 0 & 0 & 0 &  0 \\ 
  M041131 & 0 & 1 & 1 & 1 & 0 & 0 & 0 & 0 & 1 & 0 & 0 & 0 & 0 & 0 &  0 \\ 
  M041275 & 1 & 0 & 0 & 0 & 0 & 0 & 0 & 0 & 0 & 0 & 0 & 0 & 1 & 0 &  1 \\ 
  M041186 & 1 & 1 & 0 & 1 & 0 & 0 & 0 & 0 & 0 & 0 & 0 & 0 & 1 & 0 &  0 \\ 
  M041336 & 1 & 1 & 0 & 0 & 1 & 1 & 0 & 0 & 0 & 0 & 0 & 0 & 1 & 1 &  0 \\ 
  M031303 & 0 & 1 & 1 & 0 & 0 & 0 & 0 & 0 & 0 & 0 & 0 & 0 & 0 & 0 &  0 \\ 
  M031309 & 0 & 1 & 1 & 0 & 0 & 0 & 0 & 0 & 0 & 0 & 0 & 0 & 0 & 0 &  0 \\ 
  M031245 & 0 & 1 & 0 & 0 & 0 & 0 & 1 & 0 & 0 & 0 & 0 & 0 & 0 & 0 &  0 \\ 
  M031242A & 0 & 1 & 1 & 0 & 0 & 0 & 0 & 1 & 0 & 0 & 0 & 0 & 0 & 0 &  0 \\ 
  M031242B & 0 & 1 & 1 & 0 & 0 & 0 & 0 & 0 & 0 & 0 & 0 & 0 & 0 & 1 &  0 \\ 
  M031242C & 0 & 1 & 1 & 0 & 0 & 0 & 0 & 1 & 0 & 0 & 0 & 0 & 0 & 1 &  0 \\ 
  M031247 & 0 & 1 & 1 & 0 & 0 & 0 & 1 & 0 & 0 & 0 & 0 & 0 & 0 & 0 &  0 \\ 
  M031219 & 0 & 0 & 0 & 0 & 0 & 0 & 0 & 0 & 0 & 1 & 1 & 1 & 0 & 0 &  0 \\ 
  M031173 & 0 & 1 & 1 & 0 & 0 & 0 & 0 & 0 & 0 & 0 & 0 & 0 & 0 & 0 &  0 \\ 
  M031085 & 0 & 0 & 0 & 0 & 0 & 0 & 0 & 0 & 0 & 1 & 0 & 0 & 0 & 0 &  0 \\ 
  M031172 & 1 & 1 & 0 & 0 & 0 & 0 & 0 & 0 & 0 & 0 & 0 & 0 & 1 & 0 &  1 \\ 
\bottomrule
\end{tabular}
\caption{Q matrix for TIMSS 2007}
\label{TIMSS}
\end{spacing}
\end{table}

The $Q$-matrix used by \cite{lee2011cognitive} is listed in Table \ref{TIMSS}. There are 15 attributes (i.e., $K$=15) belonging to three content domains: Number, Geometric Shapes and Measures and Data Display. For this data analyses, the prior hyper-parameters and  MCMC chain lengths and burn-in are listed in Table \ref{datasetting2}. 
Due to the large number of attributes, the independent and sequential Gibbs sampling methods are the only two methods used here. 

\begin{table}[htp]
\caption{Summary of analyzing conditions for TIMSS 2007}
\begin{center}
\begin{tabular}{ccccc}
\hline
&DINA&GDINA\\
\hline
$\mbox{GDINA } \bm\lambda$ prior &---&N(0,1)\\
$\mbox{DINA } \bm s,\bm g$ prior & Beta(1,1) &---\\
$\bm\delta$ &$\bm{0.01}$ &$\bm{0.01}$ \\
Chain Length&2000&4000\\
Burn-in&1000&2000\\
\hline
\end{tabular}
\end{center}
\label{datasetting2}
\end{table}%

\begin{figure}[htp]
\centering
\subfigure[DINA Guessing parameters]{
\label{T_DINA_g}
\includegraphics[width=.31\linewidth]{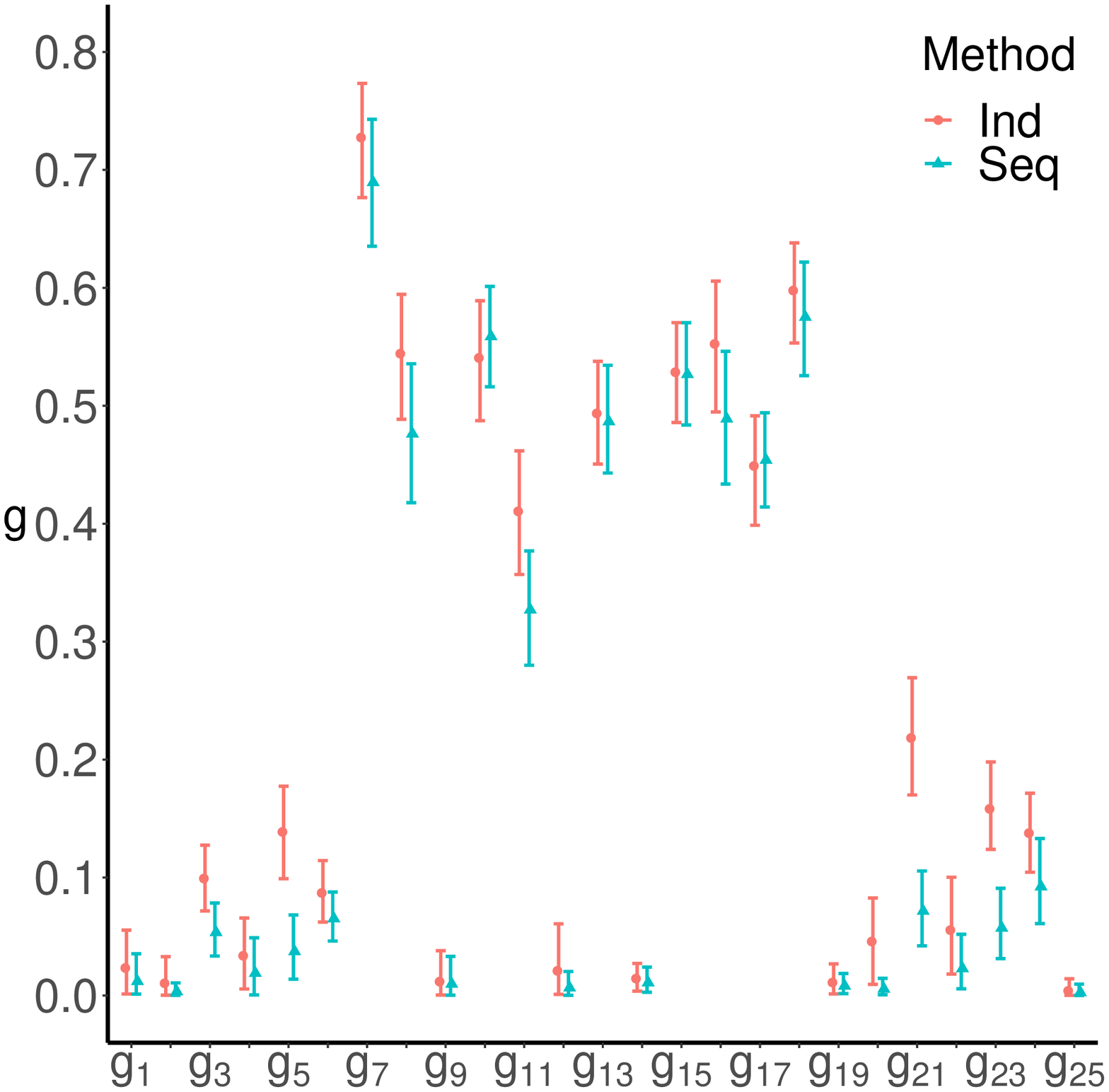}
}%
\vspace{.005\linewidth}
\subfigure[DINA Silpping parameters]{
\label{T_DINA_s}
\includegraphics[width=.31\linewidth]{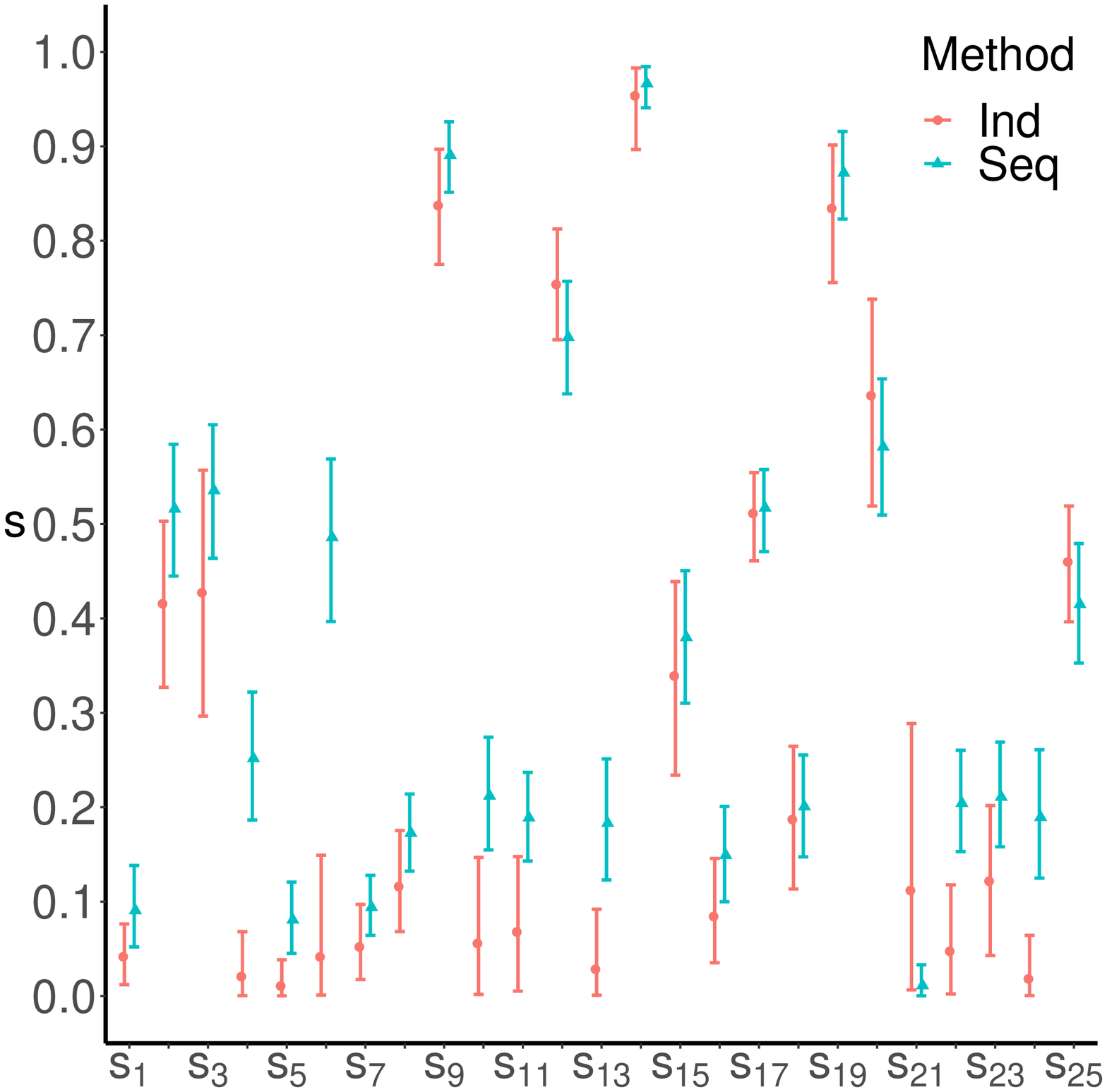}
}%
\vspace{.005\linewidth}
\subfigure[GDINA parameters]{
\label{T_GDINA}
\includegraphics[width=.31\linewidth,height =.31\linewidth]{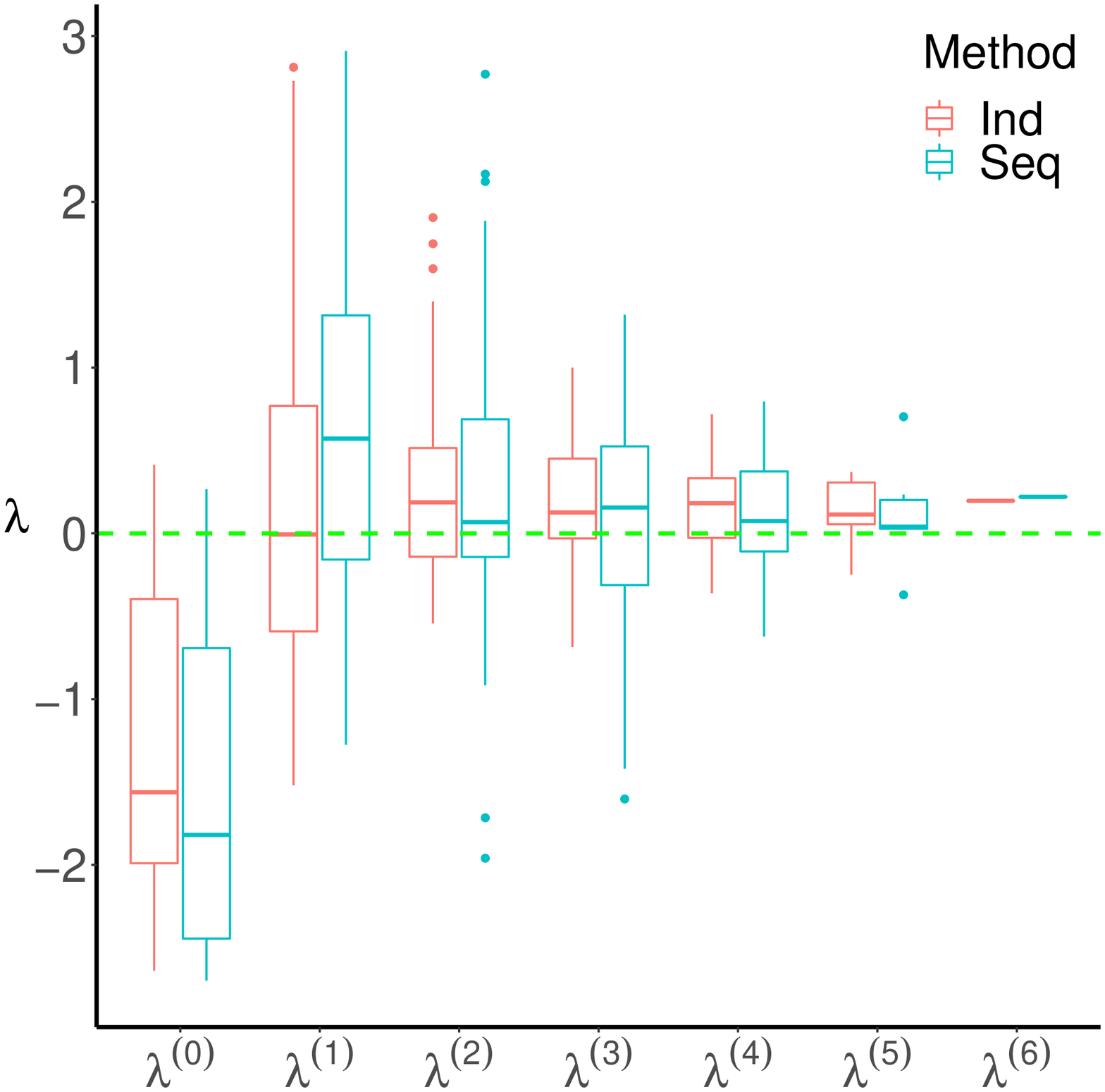}
}%
\centering
\caption{The estimations of item parameters for the TIMSS 2007 data. The ``Ind" and ``Seq” represent results from the independent and sequential Gibbs sampling, respectively.}
\label{TIMSS_DINA}
\end{figure}

Analyzing the TIMSS 2007 data with the DINA model, the time costs of the independent and sequential Gibbs sampling are are 11.51 and 13.36 seconds, respectively. The estimations and 95\% HDP regions of guessing and slipping parameters are shown in Figures \ref{T_DINA_g} and \ref{T_DINA_s}, respectively. When the DINA model is used to analyze, large guessing and slipping parameters often indicate the model may not fit the data well. 
In the results of sequential Gibbs sampling, the guessing parameters for items 15, 18, 22 and 25 are greater than 0.5, meanwhile the slipping parameters for items 2,  3,  4, 10, 11, 17, 21 and 24 are greater than 0.5. 

When to analyze the TIMSS 2007 using the GDINA model, the time costs of the independent and sequential Gibbs sampling are 35.63 and 42.75 seconds, respectively. Since the items in TIMSS 2007 need up to 6 attributes, there exist the 6-way interaction parameters. The box-plot of estimated item parameters is given in Figure \ref{T_GDINA}. We can see that the average effect of intercept is negative and the others are positive.
\end{appendix}
\end{document}